\documentstyle[11pt]{article}
\newcounter{subequation}[equation]

\makeatletter 

\def\citen#1{%
\edef\@tempa{\@ignspaftercomma,#1, \@end, }% ignore spaces in
\edef\@tempa{\expandafter\@ignendcommas\@tempa\@end}%
\if@filesw \immediate \write \@auxout {\string \citation {\@tempa}}\fi
\@tempcntb\m@ne \let\@h@ld\relax \def\@citea{}%
\@for \@citeb:=\@tempa\do {\@cmpresscites}%
\@h@ld}
\def\@ignspaftercomma#1, {\ifx\@end#1\@empty\else
   #1,\expandafter\@ignspaftercomma\fi}
\def\@ignendcommas,#1,\@end{#1}
\def\@cmpresscites{%
 \expandafter\let \expandafter\@B@citeB \csname b@\@citeb \endcsname
 \ifx\@B@citeB\relax % undefined
    \@h@ld\@citea\@tempcntb\m@ne{\bf ?}%
    \@warning {Citation `\@citeb ' on page \thepage \space undefined}%
 \else%  defined
    \@tempcnta\@tempcntb \advance\@tempcnta\@ne
    \setbox\z@\hbox\bgroup % check if citation is a number:
    \ifnum0<0\@B@citeB \relax
       \egroup \@tempcntb\@B@citeB \relax
       \else \egroup \@tempcntb\m@ne \fi
    \ifnum\@tempcnta=\@tempcntb % Number follows previous--hold on to it
       \ifx\@h@ld\relax % first pair of successives
          \edef \@h@ld{\@citea\@B@citeB }%
       \else % compressible list of successives
          \edef\@h@ld{\hbox{--}\penalty\@highpenalty
            \@B@citeB }%
       \fi
    \else   %  non-successor--dump what's held and do this one
       \@h@ld\@citea\@B@citeB
       \let\@h@ld\relax
 \fi\fi%
 \def\@citea{,\penalty\@highpenalty\hskip.13em plus.1em minus.1em}%
}
\def\@citex[#1]#2{\@cite{\citen{#2}}{#1}}%
\def\@cite#1#2{\leavevmode\unskip
  \ifnum\lastpenalty=\z@\penalty\@highpenalty\fi% highpenalty before
  \ [{\multiply\@highpenalty 3 #1%             % triple-highpenalties within
      \if@tempswa,\penalty\@highpenalty\ #2\fi % and before note.
    }]\spacefactor\@m}
\def\thesubequation{\theequation\@alph\c@subequation}
\def\@subeqnnum{{\rm (\thesubequation)}}
\def\slabel#1{\@bsphack\if@filesw {\let\thepage\relax
   \xdef\@gtempa{\write\@auxout{\string
      \newlabel{#1}{{\thesubequation}{\thepage}}}}}\@gtempa
   \if@nobreak \ifvmode\nobreak\fi\fi\fi\@esphack}
\def\subeqnarray{\stepcounter{equation}
\let\@currentlabel=\theequation\global\c@subequation\@ne
\global\@eqnswtrue
\global\@eqcnt\z@\tabskip\@centering\let\\=\@subeqncr
$$\halign to \displaywidth\bgroup\@eqnsel\hskip\@centering
  $\displaystyle\tabskip\z@{##}$&\global\@eqcnt\@ne
  \hskip 2\arraycolsep \hfil${##}$\hfil
  &\global\@eqcnt\tw@ \hskip 2\arraycolsep
  $\displaystyle\tabskip\z@{##}$\hfil
   \tabskip\@centering&\llap{##}\tabskip\z@\cr}
\def\endsubeqnarray{\@@subeqncr\egroup
                     $$\global\@ignoretrue}
\def\@subeqncr{{\ifnum0=`}\fi\@ifstar{\global\@eqpen\@M
    \@ysubeqncr}{\global\@eqpen\interdisplaylinepenalty \@ysubeqncr}}
\def\@ysubeqncr{\@ifnextchar [{\@xsubeqncr}{\@xsubeqncr[\z@]}}
\def\@xsubeqncr[#1]{\ifnum0=`{\fi}\@@subeqncr
   \noalign{\penalty\@eqpen\vskip\jot\vskip #1\relax}}
\def\@@subeqncr{\let\@tempa\relax
    \ifcase\@eqcnt \def\@tempa{& & &}\or \def\@tempa{& &}
      \else \def\@tempa{&}\fi
     \@tempa \if@eqnsw\@subeqnnum\refstepcounter{subequation}\fi
     \global\@eqnswtrue\global\@eqcnt\z@\cr}
\let\@ssubeqncr=\@subeqncr
\@namedef{subeqnarray*}{\def\@subeqncr{\nonumber\@ssubeqncr}\subeqnarray}
\@namedef{endsubeqnarray*}{\global\advance\c@equation\m@ne%
                           \nonumber\endsubeqnarray}

\@addtoreset{equation}{section}
\renewcommand{\theequation}{\thesection.\arabic{equation}}

\makeatother
\newcommand{\dalemb}[2]{{\vbox{\hrule height .#2pt
        \hbox{\vrule width.#2pt height#1pt \kern#1pt
                \vrule width.#2pt}
        \hrule height.#2pt}}}

    \let\e=\epsilon
\let\l=\lambda \let\m=\mu \let\n=\nu   

\let\L=\Lambda
 \let\Pi=\Pi \let\Sigma=\Sigma  

 \let\W=\Omega     
\def\nn{\nonumber} \def\bd{\begin{document}} \def\ed{\end{document}}
\def\ds{\documentstyle} \def\fr{\frac} \def\bl{\bigl} \def\br{\bigr}
\def\Br{\Bigr} \def\Bl{\Bigl} 
\def\bm{\bibitem}
\def\na{\nabla}
\def\pa{\partial} \def\ov{\overline}
\def\ie{{\it i.e.\ }} 
\def\etc{{\it etc.}}
\def\via{{\it via}}
\def\be{\begin{equation}} \def\ee{\end{equation}} 
\def\ba{\begin{array}} \def\ea{\end{array}} 
\newcommand{\ft}[2]{{\textstyle{{\scriptstyle #1}\over {\scriptstyle 
#2}}}} \newcommand{\fft}[2]{{#1 \over #2}} \def\del{\partial} 
\newcommand{\st}[1]{{\scriptstyle #1}} 
\newcommand{\sst}[1]{{\scriptscriptstyle #1}} \def\oneone{\rlap 
1\mkern4mu{\rm l}} \def\e7{E_{7(+7)}} \def\td{\tilde} 
\def\wtd{\widetilde} \def\im{{\rm i}} \def\bog{Bogomol'nyi\ } 
\def\q{{\tilde q}} \def\hast{{\hat\ast}} \def\0{{\sst{(0)}}} 
\def\1{{\sst{(1)}}} \def\2{{\sst{(2)}}} \def\3{{\sst{(3)}}} 
\def\4{{\sst{(4)}}} \def\5{{\sst{(5)}}} \def\6{{\sst{(6)}}} 
\def\7{{\sst{(7)}}} \def\8{{\sst{(8)}}} \def\9{{\sst{(9)}}} 
\def\ten{{\sst{(10)}}} \def\n{{\sst{(n)}}} \def\m1{{\sst{(-1)}}} 
\def\tV{{\wtd V}} \def\cF{{\cal F}} \def\cA{{\cal A}} \def\cB{{\cal 
B}} \def\hA{{\hat{\cal A}}} \def\td{\tilde} \def\wtd{\widetilde} 
\def\ep{\epsilon} \def\Z{\rlap{\sf Z}\mkern3mu{\sf Z}}
\def\R{\rlap{\rm I}\mkern3mu{\rm R}}
\def\I{{\sst I}}
\def\J{{\sst J}}
\def\K{{\sst K}}
\def\cM{{\cal M}}
\def\T{{\rm T}}
\def\v{{\cal V}}
\def\vp{{\varphi}}
\def\sA{{\sst A}}
\def\sB{{\sst B}}
\def\sC{{\sst C}}
\def\sD{{\sst D}}
\def\wb#1{{{\bar #1}}}
\def\ns{{\sst{\rm NS}}}
\def\rr{{\sst{\rm RR}}}
\def\L{{\Lambda}}

\def\cramp{\medmuskip = 2mu plus 1mu minus 2mu}
\def\cramper{\medmuskip = 2mu plus 1mu minus 2mu}
\def\crampest{\medmuskip = 1mu plus 1mu minus 1mu}
\def\uncramp{\medmuskip = 4mu plus 2mu minus 4mu}

\newcommand{\ho}[1]{$\, ^{#1}$}
\newcommand{\hoch}[1]{$\, ^{#1}$}
\newcommand{\bea}{\begin{eqnarray}} 
\newcommand{\eea}{\end{eqnarray}}
\newcommand{\bsea}{\begin{subeqnarray}}
\newcommand{\esea}{\end{subeqnarray}}
\newcommand{\ra}{\rightarrow}
\newcommand{\lra}{\longrightarrow}
\newcommand{\Lra}{\Leftrightarrow}
\newcommand{\ap}{\alpha^\prime}
\newcommand{\bp}{\tilde \beta^\prime}
\newcommand{\tr}{{\rm tr} }
\newcommand{\Tr}{{\rm Tr} } 
\newcommand{\NP}{Nucl. Phys. }
\newcommand{\tamphys}{\it Center for Theoretical Physics,
Texas A\&M University, College Station, Texas 77843}

\newcommand{\auth}{I.V. Lavrinenko\hoch{\dagger},
H. L\"u\hoch{\ddagger1}, C.N. Pope\hoch{\dagger2} and
K.S. Stelle\hoch{\star}}

\begin{document}
\begin{flushright}
CTP TAMU-09/99 \ \ \   
UPR/838-T \ \ \ 
Imperial/TP/98-99/43 \ \ \ \ \ \ 
\hfill{\bf hep-th/9903057}\\
\hfill{March 1999}\\
\end{flushright}
\begin{center}
{\Large {\bf Superdualities, Brane Tensions and Massive IIA/IIB Duality}}

\vspace{5pt}

\auth

\vspace{5pt}
{\hoch{\dagger}\tamphys}

\vspace{5pt}
{\hoch{\ddagger}\it Dept. of Phys. and Astro., University of Pennsylvania,
Philadelphia, PA 19104}

\vspace{5pt}
{\hoch{\star} \it The Blackett Laboratory, Imperial College\\
Prince Consort Road, London SW7 2BZ, UK}

\vspace{5pt}

\underline{ABSTRACT}
\end{center}

     The gauge transformations of $p$-form fields in supergravity
theories acquire a non-commuting character when one introduces
potentials both for the theory's original field strengths and for
their duals.  This has previously been shown in the ``doubled''
formalism for maximal supergravities, where a generalised duality
relation between original and dual field strengths replaces the
equations of motion.  In the doubled formalism, the gauge
transformations generate a superalgebra, and the corresponding
symmetries have accordingly been called ``superdualities.''  The
corresponding Noether charges form a representation of the cohomology
ring on the spacetime manifold.  In this paper, we show that the gauge
symmetry superalgebra implies certain non-trivial relations among the
various $p$-brane tensions, which can straightforwardly be read off
from the superalgebra commutation relations.  This provides an elegant
derivation of the brane-tension relations purely within a given
theory, without the need to make use of duality relations between
different theories, such as the type IIA/IIB T-duality, although the
results are consistent with such dualities.  We present the complete
set of brane-tension relations in M-theory, in the type IIA and type
IIB theories, and in all the lower-dimensional maximal supergravities.
We also construct a doubled formalism for massive type IIA
supergravity, and this enables us to obtain the brane-tension
relations involving the D8-brane, purely within the framework of the
massive IIA theory.  We also obtain explicit transformations for the
nine-dimensional T-duality between the massive type IIA theory and the
Scherk-Schwarz reduced type IIB theory.
 
{\vfill\leftline{}\vfill
\vskip  10pt

\footnoterule
{\footnotesize \hoch{1} Research supported in part by DOE
grant DE-FG02-95ER40893 \vskip -12pt} \vskip 14pt
{\footnotesize 
        \hoch{2}        Research supported in part by DOE 
grant DE-FG03-95ER40917. \vskip -12pt}  \vskip  14pt}
\pagebreak

\setcounter{page}{1}

\tableofcontents
\addtocontents{toc}{\protect\setcounter{tocdepth}{2}}
\newpage

\section{Introduction\label{sec:intro}}

   A new formulation of the equations of motion in maximal
supergravities was recently developed in \cite{cjlp2}, in which every
field in the theory, with the exception of gravity itself, is
augmented by a ``double'' field of the dual degree.\footnote{Doubling of gauge
fields has also been considered in the context of $D=11$ supergravity, where
there is a formalism incorporating both a 3-form and a 6-form gauge potential
\cite{padua}.} Thus in general in $D$ dimensions each potential of degree
$n$ is augmented by its double, of degree $D-n-2$.  In this approach the
doubling is performed even on the dilatons and all other scalar fields,
corresponding to
$n=0$.  The effect of this doubling is that, with the exception of the
Einstein equation, all the other bosonic equations of motion are
recast into a first-order form.  In fact, as was shown in
\cite{cjlp2}, they can all be recast in the form of an algebraic
condition on a single generalised field strength that is subject to a
``twisted self-duality'' condition.

    One of the intriguing features of the doubled system is that when
one looks at the extended set of gauge transformations for the entire
set of fields, one encounters non-commutativities that were not seen
in the analogous gauge transformations for the original system of
fields.  By associating Lie algebra generators with each field in the
extended system, one can thus construct an associated symmetry
algebra.  Interestingly enough, since the generators associated with
forms of odd degrees must themselves be odd (\ie fermionic), one
generally finds that the algebra encoding the gauge symmetry
transformations is a Lie superalgebra \cite{cjlp2}.  (The only
exception to this among the maximal supergravities is the case of the type
IIB theory in $D=10$, for which all the generators are bosonic.)
Formulating the system of bosonic field equations as a twisted
self-duality condition is achieved by exponentiating the superalgebra
generators, with the various gauge potentials as parameters, and
constructing a generalised field strength ${\cal G} = d\v\, \v^{-1}$.
The twisted self-duality condition is then expressed as ${*{\cal G}} =
{\cal S}\, {\cal G}$, where ${\cal S}$ is a pseudo-involution operator
that maps between the generators of the original fields and their
doubles \cite{cjlp2}.
    
     In fact, as we shall discuss in this paper, a careful inspection
shows that the non-commutativity of certain gauge transformations can
already be seen in the framework of the canonical formalism even
before the introduction of the dual gauge potentials.  This arises
when one considers the integrated Noether charges as generators of
canonical transformations.  Non-vanishing Noether charges for local
symmetries occur only for ``large'' gauge transformations,
corresponding to cohomologically nontrivial $p$-form gauge parameters.
As a consequence, one finds that the Poisson bracket algebra of the
integrated charges gives a representation of the cohomology ring on
the underlying spacetime manifold.

     The non-commutativity of the gauge transformations allows one to
establish a set of relations among the various $p$-brane ``tensions''
(which are perhaps better thought of as the units of the corresponding
electric charges).  By using the superalgebra in the doubled
formalism, one can straightforwardly arrive at relations between the
$p$-brane tensions that could previously be derived only using rather
intricate arguments based on duality transformations and various
D-brane techniques
\cite{dlm,schw1,alwis,schw2,schw3,brax,harm,con,jxlu}.  Some of the
relations that we shall present in this paper have appeared previously
in the literature
\cite{dlm,schw1,alwis,schw2,schw3,brax,harm,con,jxlu}, but many are
new.

     In section 2 we shall present a canonical discussion of the gauge
transformations, and shall show how the Poisson brackets of the gauge
generators can be non-vanishing even when the gauge transformations
might ostensibly appear to be abelian.  The explanation for this
apparent discrepancy is related to the subtle distinction between
gauge transformations with exact gauge parameters and transformations
with closed gauge parameters.  In fact, precisely the same subtlety
was shown in \cite{cjlp2} to be responsible for the non-commutativity
in the doubled formalism.  In section 3 we shall exploit this
non-commutativity in order to derive relations between the tensions
for the various $p$-branes supported by the fields in the
eleven-dimensional and ten-dimensional maximal supergravity theories.
As we shall show, there is in general a one-to-one correspondence
between the set of non-vanishing (anti)commutators in the Lie
superalgebra, and the set of brane-tension relations.  In section 4 we
shall extend this discussion to all the lower-dimensional maximal
supergravities.  One interesting feature is that certain sets of
brane-tension relations are themselves inter-related, as a consequence
of a discrete set of relations among the various non-trivial
commutators in the Lie superalgebras.  This application of the
so-called ``jade rule''\footnote{The term ``jade rule'' was a more
lapidary variant of general rules such as the golden rule, \etc}
\cite{cjlp2} of the Lie superalgebras leads to a significant
simplification of the structures of the brane-tension relations in the
theories.

    The brane-tension relations in the various dimensions can be
inter-related also by means of dimensional reduction, and also by
exploiting the T-duality symmetry that relates the type IIA and type
IIB theories.  We shall discuss this is detail in sections 4 and 5.
In order to obtain a complete picture, it is necessary to extend the
discussion of the type IIA theory to include the massive IIA
supergravity first constructed in \cite{romans}.  The main topic
covered in section 5 is the construction of the doubled system of
equations for the massive IIA theory, yielding an extended Lie
superalgebra with additional (anti)commutators related to tension
relations involving the D8-brane.  Finally, in an extensive appendix,
we derive explicit results for the T-duality between the massive IIA
and the type IIB supergravities.  This involves performing a
Kaluza-Klein reduction of the massive IIA theory to $D=9$, and a
generalised Scherk-Schwarz reduction of the type IIB theory to $D=9$.
We do this at the level of the full doubled systems.  In the last
subsection of the appendix, we derive the explicit field
transformations that map between the nine-dimensional massive IIA and
IIB theories.

\section{Local symmetry Noether charges and non-commutativity of 
supergravity gauge transformations}

     Let us begin with an elementary discussion of the gauge
transformations in supergravity theories and their non-commutativity.
We shall consider this issue both at the level of the gauge
transformations themselves and also at the level of the corresponding
charges.  For this, we shall first need to consider the nature of the
Noether charges that can be associated to gauge symmetries.

     It is well known that if the Lagrangian of a theory is left
invariant by some set of group transformations, one can always,
following the Noether procedure, define a set of locally-conserved
quantities, \ie Noether currents.  The conservation law for the
Noether current follows from the equations of motion.  This
conservation law has the consequence that if one integrates the time
component of the current over the volume of a spatial hypersurface,
one obtains a globally-conserved quantity, \ie a charge.

     In the case of a rigid symmetry transformation, a Noether charge
may be interpreted as the generator of the associated symmetry
transformation.  In the case of a gauge symmetry, on the other hand,
the equations of motion typically imply that the Noether charge
reduces to a surface integral.  This surface integral can sometimes be
interpreted as the generator of a non-vanishing symmetry
transformation, depending on the topological character of the
corresponding gauge parameter $\Lambda$.  Consequently, in discussing
the charges, one must take care to consider the topological character
of the corresponding gauge parameter $\Lambda$.  Thus, instead of just
considering a charge $Q$ for a given symmetry, one should consider the
charge $Q_{\Lambda}$ associated to a specific gauge transformation,
incorporating the transformation parameter into the charge integral.
For ``little'' local symmetry transformations, which can be
continuously deformed back to the identity transformation and which
fall off sufficiently fast at infinity, the total charge integral
vanishes upon use of the equations of motion.  For topologically
nontrivial, or ``large'' symmetry transformations, on the other hand,
this integral need not vanish.  In that case, a nonvanishing
integrated charge may be interpreted, {\via} Poisson (or, more
correctly, Dirac) brackets, as the generator of a gauge
transformation.  Since ``little'' gauge transformations have vanishing
charge integrals, there is a natural equivalence relation between
large symmetry transformations differing by little transformations.
In view of this behaviour, the large symmetry transformations are
somewhat akin to rigid symmetry transformations such as Yang-Mills
colour-rotating transformations that tend to constants instead of
falling off at infinity, for which nonvanishing Noether charges may be
defined, and for which charge integrals corresponding to
transformations differing by a ``little'' gauge transformation are
equal.

     Let us now consider the construction of Noether currents and
charges more specifically. If we have a set of fields $\phi^i$, where
$i$ labels the fields, and a set of transformations
$\delta\phi^i=f^i(\phi^j)$ which leave the Lagrangian ${\cal L}$
invariant , where $f^i(\phi^j)$ are some given functions, then the
conserved Noether current is given by
%%%%%
\be
j^{\mu}=\fft{\pa {\cal L}}{\pa\pa_{\mu}\phi^i} \delta\phi^i
\ .\label{curr}
\ee
%%%%%
This definition is appropriate in the case where Lagrangian itself is
invariant under the symmetry transformations.  If instead it is
invariant only up to a total derivative, \ie if it transforms as
$\delta{\cal L}=\pa_{\mu}\W^{\mu}$ for some $\W^{\mu}$, then the
formula (\ref{curr}) is replaced by
%%%%%
\be
j^{\mu}=\fft{\pa {\cal L}}{\pa\pa_{\mu}\phi^i} \delta\phi^i-\W^{\mu}\ .
\label{curr1}
\ee
%%%%%

     This last expression is the one that we shall be using, since it
is indeed the case that some gauge transformations leave the
supergravity Lagrangians invariant only up to total derivatives.
Having established the notation, we shall now derive explicit
commutation relations in the simplest of the examples, namely
eleven-dimensional supergravity.  Let us first note a simplifying
feature of the formula (\ref{curr1}).  Since we are interested in
commutation relations for globally-conserved charges only, we have to
consider the following integral
%%%%%
\be
Q=\int j^0 dV^{(10)},
\ee
%%%%%
where integration is performed over the entire ten-dimensional
space. Note that in the definition of $j^0$
(\ref{curr1}), the first term is nothing but the canonical momentum
multiplied by the field variation under the symmetry transformation.

    The field content of eleven-dimensional supergravity includes a 3-form
$A_\3$. It is a gauge field, transforming as $\delta A_\3=\L_\3$ under gauge
transformations\footnote{This version of the $p$-form gauge transformation,
which allows for cohomologically nontrivial gauge parameters, has been discussed
in particular in Ref.\ \cite{juliasilva}.} where $\L_\3$ is an arbitrary {\it
closed} 3-form, $d\L_\3=0$. It is a straightforward calculation to see that the
eleven-dimensional Lagrangian
%%%%%
\be
{\cal L}_{11}=R*1-\ft12*F_\4\wedge F_\4-
\ft16 F_\4\wedge F_\4\wedge A_\3 \label{11lagr}
\ee
%%%%%
transforms as 
%%%%%
\be
\delta {\cal L}=d(\ft16\L_\3\wedge A_\3\wedge F_\4)\ ,\label{lagtr}
\ee
%%%%%
which implies, according to our previous discussion, that the
following conserved charges can be defined:
%%%%%
\be
Q_e(\L_\3)=\int\L_\3\wedge (*\Pi-\ft16 A_\3\wedge F_\4)\ .  
\label{Q3char} 
\ee
%%%%%
It is understood that the integrand here is projected into a
10-dimensional spacelike hypersurface.  In (\ref{Q3char}), we have
introduced a canonical momentum 3-form $\Pi=\fft{1}{3!}\Pi_{ijk}
dx^i\wedge dx^j\wedge dx^k$, with components defined by
$\Pi_{ijk}=\fft{\pa{\cal L}}{\pa(\pa_t A_{ijk})}$.  The Hodge dual is
taken with respect to the ten-dimensional metric.  One may verify that
for a $\L_\3$ that is not only closed but also exact, and vanishing
sufficiently fast at infinity, the charge integral (\ref{Q3char}) vanishes upon
integration by parts and the use of the equations of motion.  This behaviour may
be compared with the analogous charge integral in Maxwell theory,
$\int\L_\1\wedge *\Pi$, where
$\Pi^i=F^{0i}$.  For an exact $\L_\1$ falling off sufficiently rapidly at
infinity, this integral vanishes upon use of the equations of motion.
But for non-exact $\L_\1$, the integral need not vanish.

     For the charges (\ref{Q3char}), one may use the canonical Poisson
bracket relations
%%%%%
\be
\{A_{\i_1\i_2\i_3}, {\Pi'}^{\j_1\j_2\j_3}\}=
6\delta_{[\i_1}^{[\j_1}\, \delta_{\i_2}^{\j_2}\, \delta_{\i_3]}^{\j_3]}\,
\delta^{(10)}(x-x')\ ,
\ee
%%%%%
with all others vanishing, to derive the charge algebra
%%%%%
\be
\{Q_e(\L^1_\3), Q_e(\L^2_\3)\}=Q_m(\L^1_\3\wedge \L^2_\3)\ 
,\label{chargering} \ee
%%%%%
where the charge $Q_m(\L_\6)$ is defined by
%%%%%
\be
Q_m(\L_\6)=\int \L_\6\wedge F_\4\ .
\ee
%%%%%
With this result we see that the non-commutativity is a characteristic
property of the theory, rooted in the structure of the gauge symmetry,
and is not just an incidental by-product of the doubled formalism that
we are using.  In the doubled formalism, on the other hand, we shall
encounter in addition an underlying (super)algebra of gauge
transformations that accords with the algebra (\ref{chargering}) for
the integrated charges.

     Similar discussions can be given in other situations where we
meet non-commutativity of form-field gauge transformations, for
example in the type IIA and IIB theories.  The algebras in those cases
are a little more complicated, but the basic structure remains the
same.

     At the level of the integrated charges, the algebra
(\ref{chargering}) reflects the ring structure of the cohomology of
$p$-form gauge parameters on the underlying spacetime manifold.  Thus,
another interpretation of the integrated charges such as
(\ref{Q3char}) is as a representation of the cohomology ring of the
spacetime manifold.  For most of the examples that arise in the study
of $p$-brane solutions in supergravity, this cohomology ring
corresponds to that of a torus.  It remains an interesting problem to
explore more sophisticated situations with manifolds of less trivial
cohomology.

\section{Brane tension relations}

    In this section we shall derive some direct consequences of the
non-commutativity of certain gauge transformations in supergravity
theories. It turns out that consistency requirements impose some
rather nontrivial relations among the various $p$-brane tensions. Some
of the relations we derive here have appeared previously in the
literature \cite{dlm,schw1,alwis,schw2,schw3,brax,harm,con,jxlu},
where they were obtained by more indirect means.  Typically, this
involved making a sequence of mappings between different low-energy
theories, for example by exploiting the T-duality that relates the
type IIA and IIB theories, or even more indirectly through the
requirements for certain anomaly cancellations. By contrast, the
method that we shall present below represents a considerable
simplification, not only technically but also conceptually, in that it
allows the brane-tension relations to be derived purely within the
framework of the low-energy description of a given theory.

\subsection{M-brane tensions}

    Let us start with the simplest example, namely eleven-dimensional
supergravity. We shall be rather brief, since the doubled formalism
has been developed in detail in an earlier paper \cite{cjlp2}; we
refer the interested reader there for additional information.  The
bosonic Lagrangian for eleven-dimensional supergravity is given by
(\ref{11lagr}).  Varying with respect to $A_\3$, we obtain the
equation of motion
%%%%%
\be
d{*F_\4}+\ft12 F_\4\wedge F_\4=0\ .\label{eqF}
\ee
%%%%%
Equation (\ref{eqF}) can be written as $d({*F_\4} +\ft12 A_\3\wedge
F_\4)=0$, and so we can write the field equation in the first-order
form
%%%%%
\be
*F_\4=F_\7\equiv dA_\6-\ft12 A_\3\wedge F_\4\ , \label{dual}
\ee
%%%%%
where we have introduced the dual potential $A_\6$. It is easy to
check that the first-order equation (\ref{dual}) is invariant under
the following gauge transformations \cite{cjlp2}
%%%%%
\be
\delta A_\3=\L_\3,\qquad \delta A_\6=\L_\6-\ft12 \L_\3\wedge A_\3\ ,
\label{gsym}
\ee
%%%%%
where $\L_\3$ and $\L_\6$ are closed 3-form and 6-form gauge
parameters, satisfying $d\L_\3=0$ and $d\L_\6=0$. The commutators of
infinitesimal gauge transformations are given by
%%%%%
\bea 
{[}\delta_{\L_\3}, \delta_{{\L'}_\3}{]} &=& \delta_{{\L''}_\6},\qquad 
{\L''}_\6 = \L_\3 \wedge {\L'}_\3,\nn\\ 
{[}\delta_{\L_\3}, \delta_{\L_\6}{]} &=& 0, 
\qquad {[}\delta_{\L_\6}, \delta_{{\L'}_\6}{]} = 0\ .\label{comm}
\eea
%%%%%
Since these transformations are to be thought of as gauge symmetries,
it follows that not only the eleven-dimensional equations of motion
for the massless fields, but also the low-energy actions for all
extended objects, including massive $p$-branes, must be invariant
under these symmetries. In order to investigate the restrictions
imposed by their non-commutativity, we need to incorporate the
couplings of the $(p+1)$-forms to the world-volume fields.
Fortunately, the nature of these couplings is well known.  For
example, the term in the world-volume action describing the coupling
of the 3-form $A_\3$ has the minimal form
%%%%%
\be
T_\3\, \int A_\3,\label{coup}
\ee
%%%%%
where $T_\3$ is the membrane ``tension.''  In this paper, we shall use
the letter $T$ exclusively for brane tensions.  In particular
$T_{\sst{(d)}}$, with $d=p+1$ denotes the tension for the $p$-brane.

    Let us suppose now that the space-time contains a compact
six-dimensional sub-manifold with non-trivial third and sixth homology
groups, $H^3(M)$ and $H^6(M)$.  To simplify the discussion, we shall
take this compact sub-manifold to be the six-torus. Now, if we wrap a
membrane around one of the homology three-cycles $M_\3$, and consider
making gauge transformations of the form $\delta A_\3=\omega_\3$,
where $\omega_\3$ is a closed 3-form such that
$\int_{M_\3}\omega_\3\neq 0$, then invariance of the term (\ref{coup})
imposes the following restriction on the value of the integral
%%%%%
\be
T_\3\, \int_{M_\3}\omega_\3=2\pi k,\label{quant}
\ee
%%%%%
where $k$ is an arbitrary integer, in order that the quantum effective
action be invariant.  This condition implies that if we take any
closed 3-form gauge parameter, and integrate it over an arbitrary homology
3-cycle, then result must be quantised in terms of the inverse membrane
tension.

     From our above discussion we know that the gauge transformations
do not commute.  However, since the commutator of two symmetry
transformations must itself also be a symmetry, we conclude that it
too should leave everything invariant.  We have seen above that the
commutator of two gauge transformations of the potential $A_\3$ gives
rise to a gauge transformation of the potential $A_\6$ (\ref{comm}).
It is natural to think of $A_\6$ as the gauge potential for the
magnetic field-strength, and as such it must couple minimally to the
world-volume of the five-brane through the term
%%%%%
\be
T_\6\, \int A_\6\ .\label{coup1}
\ee
%%%%%
The $A_\6$ potential has its own independent symmetry; namely, we can shift it
by an arbitrary closed 6-form, $\omega_\6$.  By the same argument as for the
membrane, provided that five-brane is also wrapped over a certain homology
6-cycle the invariance of the world-volume action implies the quantisation
condition
%%%%%
\be
T_\6\int_{M_\6}\omega_\6=2\pi \ell\ ,\label{quant1}
\ee
%%%%%
where $\ell$ is again an arbitrary integer. Now, if we commute two
gauge transformations for the $A_\3$ potential we obtain the following
shift in the five-brane world-volume action:
%%%%%
\be
\delta {\cal S}=T_\6\, \int_{M_\6}\omega^1_\3\wedge\omega^2_\3,
\ee
%%%%%
where $\omega^1_\3$ and $\omega^2_\3$ are the parameters of the first
and second gauge transformations respectively. For the torus this
integral can be decomposed into the sum of products of integrals over
3-cycles. But we already know that these integrals are quantised in
terms of the membrane tension (\ref{quant}).  This is consistent with
the equation (\ref{quant1}) if and only if
%%%%%
\be
T_\6=\fft{1}{2\pi}\, T_\3^2\ .\label{d11tension}
\ee
%%%%%

      At this stage it is worthwhile to make an observation that
significantly simplifies the calculations in more complicated cases,
such as the type IIA or IIB theories in ten or lower dimensions.  One
can recast the commutation relations (\ref{comm}) as commutators in an
ordinary super Lie algebra, by introducing generators $V$ and $\td V$
for the $\L_\3$ and $\L_\6$ transformations respectively.  We see that
commutation relations (\ref{comm}) translate into the super Lie
algebra \cite{cjlp2}
%%%%%
\be
\{V, V\}=-\td V, \qquad {[}V, \td V{]}=0, \qquad {[}\td V, \td V{]}=0.
\label{alg}
\ee 
%%%%%
Note that commutators are even or odd according to whether the degrees
of the associated field strengths are odd or even.  (We shall in
general, unless severe ambiguity might arise, avoid clumsy language by
referring to commutators and anti-commutators generally as commutators
in what follows.)  Again we refer the reader to \cite{cjlp2} for all
details about these algebras.  Here, we wish only to point out that
the complete structure of all relations among the $p$-brane tensions
is encoded in these algebras, and can be directly read off from the
commutators of the generators. For each non-vanishing commutator, one
simply needs to replace the bracket on the left by the product of
corresponding inverse $p$-brane tensions, each multiplied by $2\pi$,
and likewise with each term on the right.  For example, if one takes
the first commutator in (\ref{alg}), and replaces $\{V, V\}$ by
$(2\pi/T_\3)^2$ and $\td V$ by $(2\pi/T_\6)$ (minus signs must be
dropped), then one arrives at the relation (\ref{d11tension}).

\subsection{Type IIA brane tensions}

     The gauge potentials in type IIA massless supergravity in $D=10$
are $A_\3$, $A_{\2 1}$, arising from the dimensional reduction of the
three-form potential in $D=11$, together with the Kaluza-Klein vector
$\cA_\1^1$.  (Note that the index 1 implies that it is the first step
in the reduction, from $D=11$ to $D=10$.) Their dual potentials are
$\wtd A_\5$, $\wtd A^1_\6$ and $\wtd A_{\7 1}$ respectively.  These
fields, together with the dilaton $\phi$ and its 8-form dual $\psi$
can be used to construct a ``coset representative'' as follows:
%%%%%
\be {\cal V} = e^{\fft12\phi\, H}\, e^{{\cal A}^1_\1\, W_1}\, 
e^{A_{\2 1}\, V^1}\, e^{A_\3\, V}\, e^{\td A_\5\, \tV}\, 
e^{\td A^1_\6\, \tV_1}\, e^{\wtd{\cal A}_{\7 1}\, \wtd W^1}\, 
e^{\fft12\psi\, \td H}\ .
\ee
%%%%%
Here, the generators $H$, $W_1$, $V^1$, $V$, $\wtd V$, $\wtd V_1$,
$\wtd W^1$ and $\wtd H$ satisfy the following super Lie algebra
\cite{cjlp2}
%%%%%
\bea
{[} H, W_1 {]} &=& \ft32 W_1\ ,\qquad
{[} H, V^1 {]} = -V^1\ ,\qquad
{[} H, V {]} = \ft12 V\ ,\nn\\
{[} H, \wtd W_1 {]} &=& -\ft32 \wtd W_1\ ,\qquad
{[} H, \wtd V^1 {]} = \wtd V^1\ ,\qquad
{[} H, \wtd V {]} = -\ft12 \wtd V\ ,\nn\\
{[}W_1, V^1{]} &=& -V\ ,\qquad \{W_1,\td V\} = -\wtd V_1\ ,\qquad
{[}V^1,V{]} = - \wtd V\ ,\nn\\
{[} V^1,\wtd V{]} &=& -\wtd W^1\ ,\qquad \{V,V\} = -\wtd V_1\ ,\qquad
\{W_1, \wtd W^1\}= -\ft38 \wtd H\ ,\nn\\
{[}V^1, \wtd V_1{]} &=& -\ft14 \wtd H\ ,\qquad 
\{ V,\wtd V\} = -\ft18 \wtd H\ ,\label{2athcom}
\eea
%%%%%
with all other commutators vanishing.  The equations of motion are
then given by ${*\cal G}= {\cal S}\, {\cal G}$, where $*$ is the Hodge
dual and ${\cal G}=d\v\, \v^{-1}$.  The operator ${\cal S}$ is an
involution (or, according to circumstance, a pseudo-involution) that
exchanges each generators for a field with that of its partners under
the doubling \cite{cjlp2}.

It is now a rather straightforward procedure to read off a variety of
relations among all the $p$-brane tensions in the type IIA theory.
For instance, the brackets involving $W_1$ or $\wtd W^1$, associated
with the Kaluza-Klein vector and its dual, give rise to the following
identities
%%%%%
\be
T_\7=\fft{1}{2\pi}\,T_\2 \, T_\5\ ,\quad
T_\6=\fft{1}{2\pi}\,T_\1\, T_\5\ ,\quad
T_\3 = \fft{1}{2\pi}\, T_\1\, T_\2\ .\label{d10tension1}
\ee
%%%%%
The brackets involving only $V$'s and $\wtd V$'s, associated with the
fields coming from the dimensional reduction of $A_\3$ and its dual 
in $D=11$, give rise to
%%%%%%%%
\be
T_\6 = \fft{1}{2\pi}\, T_\3^2\ ,\qquad
T_\5 = \fft{1}{2\pi}\, T_\2\, T_\3\ .\label{d10tension2} 
\ee
%%%%%
Note that the first relation in (\ref{d10tension2}) is the same as the
one found already in $D=11$, and hence it can be viewed as a vertical
dimensional reduction of the result in $D=11$.  The second relation in
(\ref{d10tension2}) can be viewed as a double-dimensional reduction
from (\ref{d11tension}) in $D=11$.  The relations in
(\ref{d10tension1}) involve the tensions of the D0-brane and D6-brane,
which are associated with the Kaluza-Klein vector, and hence they are
not related to (\ref{d11tension}) by dimensional reduction.  Note that
there is a conservation rule for the subscripts that denote the
word-volume dimensions of the various branes appearing in the tension
relations.

     It is worth mentioning that all the tensions in this paper are
measured using the $p$-branes' own metrics, {\it e.g.}\ string tension
is measured in the string metric; membrane tension is measured in the
membrane metric, \etc.  In such metrics, the tensions are independent
of the moduli.  One can of course also discuss the tensions in a given
fixed metric.  In that case, the tensions would in general depend on
the moduli, since the metrics are related by modulus-dependent Weyl
transformations.  It is straightforward to generalise to these cases,
following from the fact that if we have an algebra $[X, Y\} = Z$, then
we have a dilaton summation rule that the dilaton vector coupled to
the field associated with the generator Z is the sum of the dilaton
vectors of the fields associated with $X$ and $Y$.  This dilaton
summation rule guarantees the proper dilaton dependence in the brane
tension relations in any given metric.

     We should draw attention to a subtlety in the use of such
algebras for extracting relations among $p$-brane tensions. In the
super Lie algebra, there are generators $H$ and $\wtd H$ associated
with the dilaton and its dual $\psi$.  However, there seem to be no
BPS objects in the supergravity theories that naturally couple to a
dilaton.  It follows that commutators involving $H$ and $\wtd H$ do
not imply any tension relations.  Furthermore, only the non-vanishing
commutators of generators associated with gauge potentials (which can
include axions) are associated with non-trivial tension relations
among the corresponding $p$-branes.

\subsection{Type IIB brane tensions}

     The doubled formalism for type IIB theory can be constructed by
introducing the dual potentials $\psi$, $\wtd \chi$, $A_\6^i$ for the
original fields $\phi$, $\chi$, and $A_\2^i$.  Note that of the index
values $i=1,2$, the value $i=1$ corresponds to NS-NS fields, while
$i=2$ corresponds to R-R fields.  Introducing a generator for each
potential as before, one can construct a coset representative
\cite{cjlp2}
%%%%%
\be
{\cal V} = e^{\fft12\phi\, H}\, e^{\chi\, E_+}\, e^{(A_\2^1\, V_+ +
A_\2^2\, V_-)}\, e^{B_\4\, U}\, e^{(A_\6^1\, \wtd V_+ + A_\6^2\,
\wtd V_-)}\, e^{\wtd\chi\, \wtd E_+}\, e^{\fft12\psi\, \wtd H}\ .
\label{2bcoset}
\ee
%%%%%
The equations of motion can then be written as ${*\cal G}={\cal S}\,
{\cal G}$, with ${\cal G} = d\v\, \v^{-1}$, provided that the
generators satisfy the super Lie algebra \cite{cjlp2}
%%%%%
\bea
&&{[}H, E_+{]} = 2E_+\ ,\qquad {[}H, V_+{]} = V_+\ ,\qquad
{[}H, V_-{]} = -V_-\ ,\nn\\
&&{[}H, \wtd E_+{]} = -2\wtd E_+\ ,\qquad {[}H, \wtd V_+{]} = -\wtd V_+\ ,
\qquad
{[}H, \wtd V_-{]} = \wtd V_-\ ,\nn\\
&&{[}E_+, V_-{]}= V_+\ ,\qquad {[}E_+, \wtd V_+{]}= -\wtd V_-\ ,\qquad
{[}V_+, V_-{]} = - U\ ,\nn\\
&& {[}V_+, U{]} =\wtd V_-\ ,\qquad
   {[}V_-, U{]} =-\wtd V_+\ ,\qquad
   {[}V_-, \wtd V_+{]}=\wtd E_+\ ,\nn\\
&& {[}E_+, \wtd E_+{]} = \ft12\wtd H\ ,\qquad
   {[} V_+, \wtd V_+{]} =\ft14\wtd H\ ,\qquad
   {[}V_-, \wtd V_-{]} = -\ft14\wtd H\ .\label{2bcom}
\eea
%%%%%

      Thus, this set of algebraic relations again enables us simply to
read off the relations among the type IIB $p$-brane tensions, namely
%%%%%%%
\bea
&&T^{\ns}_\6=\fft{1}{2\pi}\,T^\rr_\2\,  T_\4\ , \qquad
T^\rr_\6=\fft{1}{2\pi}\,T^{\ns}_\2 \, T_\4\ ,\qquad
T_\4=\fft{1}{2\pi}\, T^{\ns}_\2\,  T^\rr_\2\ ,\label{d102btension1}\\
&&T_\8=\fft{1}{2\pi}\, T^{\ns}_\2 \, T^\rr_\6\ ,\qquad
T^\rr_\2=\fft{1}{2\pi}\,T_{\0}\,  T^{\ns}_\2\ ,\qquad
T^{\ns}_\6=\fft{1}{2\pi}\, T_{\0}\,  T^\rr_\6\ .\label{d102btension2}
\eea
%%%%%
Note that the tension relation (\ref{d102btension1}) is $SL(2,\R)$
covariant, whilst (\ref{d102btension2}) is not.  This is
understandable, since the higher-degree gauge potentials form linear
representations under $SL(2,\R)$, and hence so do their associated
tensions.  The tensions $T_\0$ and $T_\8$ are associated with the
axion and its dual, which do not transform linearly under $SL(2,\R)$,
and hence (\ref{d102btension2}) is not $SL(2,\R)$ covariant.

\section{Lower-dimensional brane tensions}

        In the previous sections, we have showed that the brane
tension relations in M-theory or in the type II theories can be
derived from the non-commutativity of the gauge transformations in the
corresponding supergravities.  In particular, they can be read off
directly from the super Lie algebras of the associated doubled
formalisms constructed in \cite{cjlp2}.  The super Lie algebras for
all lower-dimensions maximal massless supergravities were also
obtained in \cite{cjlp2}, and from these it is straightforward to read
off the complete set of brane tensions in all the toroidally-reduced
theories.

\subsection{The reduction rule and the brane-tension ``jade rule''}

       We begin with a brief review of the super Lie algebra of the
lower dimensional maximal massless supergravities.  These can be
obtained by dimensional reduction from $D=11$ supergravity or type IIB
supergravity.  In the bosonic sector, in additional to the metric, the
theory contains the dilatons $\vec \phi$ and the gauge potentials
$\cA^i_{\0j}$, $\cA^i_\1$, $A_{\0ijk}$, $A_{\1 ij}$, $A_{\2 i}$ and
$A_\3$.  In the doubled formalism, a dual field is introduced for each
field (except for the metric), giving $\vec \psi$, $\wtd
\cA^i_{\sst{(D-2)} j}$, $\wtd \cA_{\sst{(D-3)} i}$, $\wtd A_{\sst
{(D-2)}}^{ijk}$, $\wtd A_{\sst{(D-3)}}^{ij}$, $\wtd
A_{\sst{(D-4)}}^{i}$ and $\wtd A_{\sst{(D-5)}}$.  The associated
generators for all these fields are given by $\vec H$, $E_i{}^j$,
$W_i$, $V^{ijk}$, $V^{ij}$, $V^i$ and $V$ for the original fields, and
$\wtd E^i{}_j$, $\wtd W^i$, $\wtd V_{ijk}$, $\wtd V_{ij}$, $\wtd
V_{i}$ and $\wtd V$ for the doubled fields.

       The generators form a deformed cotangent super Lie algebra.  To
be precise, let use $U^a$ to denote the set of generators of $\{\vec
H, E_i{}^j, W_i\}$, that is associated with the fields coming from the
dimensional reduction of the metric, and $U^{\bar a}$ to denote the
generators of $\{V^{ijk}. V^{ij}, V^{i}, V\}$ that are associated with
the fields coming from the dimensional reduction of the three-form
potential in $D=11$.  Then the superalgebra has the following form
\cite{cjlp2}
%%%%%%%
\bea
&&{[} U^a, U^b \} = f^{ab}{}_c\, U^c\ ,\qquad
  {[} U^a, U^{\bar b} \} = f^{a\bar b}{}_{\bar c}\, U^{\bar c}\ ,\qquad
  {[} U^{\bar a}, U^{\bar b} \} = g^{\bar a\bar b \bar c}\, \wtd
U_{\bar c}\ ,\nn\\
&&{[} U^a, \wtd U_b \} = f^{ca}{}_b\, \wtd U_c\ ,\qquad
  {[} U^a, \wtd U_{\bar b} \} = f^{\bar c a}{}_{\bar b}\,
\wtd U_{\bar c}\ ,\qquad
  {[} U^{\bar a}, \wtd U_{\bar b} \} = f^{c\bar a}{}_{\bar b}\, \wtd
U_c\ .\label{genalg1}
\eea
%%%%%%%%
This algebra satisfies the so-called ``jade rule'', which states that
if we have untilded generators $X$, $Y$ and $Z$ where ${[}X,Y\} = Z$,
then it follows that we will necessarily also have ${[}X,\wtd Z \} =
(-1)^{XY+1}\, \wtd Y$ \cite{cjlp2}.  This implies that once the
structure constants in the first line in (\ref{genalg1}) are given,
the structure constants for the second line can be deduced from the
jade rule.  Thus it is only necessary for us to present the
commutation relations for $U^a$ and $U^{\bar a}$, which are given by
%%%%%%%%%%%
\bea
&&{[} E_i{}^j, E_k{}^\ell {]} = \delta^j_k\, E_i{}^\ell -
\delta^\ell_i\, E_k{}^j\ ,\qquad
{[}E_i{}^j, E^{k\ell m} {]} = - 3\delta^{[k}_i\, E^{\ell m]j}\ ,\nn\\
&&{[} E_i{}^j, V^k{]} = -\delta^k_i\, V^j\ ,\qquad
  {[} E_i{}^j, V^{k\ell}{]} = 2\delta^{[k}_i\, V^{\ell] j}\ ,\qquad
   {[}E_i{}^j, W_k{]} = \delta^j_k\, W_i\ ,\nn\\
&&{[} W_i, E^{jk\ell}{]} = - 3\delta_i^{[j}\, V^{k\ell]}\ ,\qquad
\{ W_i, V^{jk} \} = -2\delta_i^{[j}\, V^{k]}\ ,\qquad
{[} W_i, V^j{]} = -\delta_i^j\, V\ ,\nn\\
&&{[} V^{\bar a}, V^{\bar b} \} = -(-1)^{{[} \bar b {]}}\,
\epsilon^{\bar c \bar a\bar b}\, \wtd V_{\bar c}\ ,
\label{genalg2}
\eea
%%%%%%%%
together with $[\vec H, X]=\vec \mu\, X$ where $\vec \mu$ is the
dilaton vector for any generator $X$.  Note that here we use generic
indices $\bar a,\bar b,\ldots$ to represent antisymmetrised sets of
$i,j,\ldots$ indices.  The symbol ${[}\bar a{]}$ denotes the number of
such $i,j,\ldots$ indices.  Appropriate $1/{[}\bar a{]}!$ combinatoric
factors are understood in summations over repeated generic indices.
It is easy to see from (\ref{genalg2}) that the algebra for the
generators $\{\vec H, E_i{}^j, W_i\}$ is $G=SL_{+}(11-D|1)$, and the
generators $\{V, V^i, V^{ij}, V^{ijk}\}$ form representations under
$G$.

        The jade rule for the algebra (\ref{genalg1}) has the
consequence that if we have a tension relation
%%%%%%%%
\be
T_{\sst{(n+m)}} =\fft{1}{2\pi}\, T_{\sst{(n)}}\,T_{\sst{(m)}}
\label{ifs}
\ee
%%%%%%%
then we must also have two further tension relations
%%%%%%
\be
T_{\sst{(D-2-n)}} = \fft{1}{2\pi}\, T_{\sst{(D-2-n-m)}}\,
T_{\sst{(m)}} \qquad {\rm and}\qquad
T_{\sst{(D-2-m)}} = \fft{1}{2\pi}\, T_{\sst{(D-2-n-m)}}\,
T_{\sst{(n)}} \ .\label{jaderule}
\ee
%%%%%%%
For example, the M-brane tension relation (\ref{d11tension}) is
invariant under this jade rule.  The full set of brane tension
relations of the type IIA theory given in (\ref{d10tension1}) and
(\ref{d10tension2}) can be obtained from applying the jade rule on the
first equations in (\ref{d10tension1}) and (\ref{d10tension2})
respectively.  The same story goes for the type IIB case, with the
complete set of tension relations given in (\ref{d102btension1}) and
(\ref{d102btension2}).

         It follows from the above discussion that the complete set of
brane-tension relations in lower dimensions is given by
%%%%%%%%%
\bea 
&&T_{\0 i}{}^j = \fft1{2\pi}\, T_{\0 i}{}^k\, T_{\0 k}{}^j\
,\qquad T_{\0}^{ijk} = \fft1{2\pi}\, T_{\0\ell}{}^i\, T_{\0}^{\ell
jk}\ ,\nn\\ 
&&T_\2^i =\fft1{2\pi}\, T_{\0 j}{}^i\, T_\2^j\ ,\qquad
T_{\1}^{ij}=\fft1{2\pi}\, T_{\0 k}{}^i\, T_\1^{jk}\ ,\qquad T_{\1 i}=
\fft1{2\pi}\, T_{\0 i}{}^j\, T_{\1 j}\ ,\nn\\ 
&& T_{\1}^{ij} =
\fft1{2\pi}\, T_{\1 k}\, T_{\0}^{ijk}\ ,\qquad T_{\2}^i =\fft1{2\pi}\,
T_{\1 j}\, T_{\1}^{ij}\ ,\quad T_{\3} = \fft1{2\pi}\, T_{\1 i}\,
T_{\2}^i\ ,\nn\\ 
&&T^{\bar a}_{\sst{(D-2-[\bar a])}} = \fft1{2\pi}\,
T^{\bar b}_{\sst{([\bar b])}}\, T^{\bar c}_{\sst{([\bar c])}}
\ ,\label{mini} 
\eea 
%%%%%%%
together with those which can be derived from the jade rule.  Here, we
are using a self-explanatory notation for labelling the brane tensions
that parallels the index labelling on the corresponding gauge
potentials listed previously. In the last equation in (\ref{mini}), it
is only when $\{\bar a, \bar b, \bar c\}$ collectively saturate the
range of the internal indices without repetition that there is a
non-trivial relation between the associated tensions.  There is no sum
over the repeated indices in (\ref{mini}); rather it meant that the
relation holds for different values of the repeated indices.

         Having obtained the complete set of the brane-tension
relations in lower dimensions, it is of interest to see how they are
related by dimensional reduction.  If in $D+1$ dimensions there is a
tension relation given (\ref{ifs}), then in $D$ dimensions, there
exist relations
%%%%%%%
\be
T_{\sst{(n+m)}} =\fft{1}{2\pi}\, T_{\sst{(n)}}\,T_{\sst{(m)}}\ ,\qquad
T_{\sst{(n+m-1)}} =\fft{1}{2\pi}\, T_{\sst{(n-1)}}\,T_{\sst{(m)}}\ ,\qquad
T_{\sst{(n+m-1)}} =\fft{1}{2\pi}\, T_{\sst{(n)}}\,T_{\sst{(m-1)}}\ .
\ee
The first relation can be viewed as coming from vertical
dimensional reduction, whilst the
second and third come from diagonal reduction.
Of course, additional brane tensions emerge from the introduction of
a new Kaluza-Klein vector, associated with the generator $W_{i}$, whose
algebra is given in (\ref{genalg2}).

\subsection{IIA/IIB T-duality}
 
     The standard dimensional reduction of the type IIA and type IIB
supergravities on a circle gives rise to two $D=9$ supergravities
which are identical, modulo field redefinitions.  The identification
of the type IIA/IIB gauge potentials leads to an identification of
their associated electric and magnetic brane tensions.  It is
straightforward then to see that the $D=9$ brane tensions relations
are the same in the two theories obtained from standard dimensional
reduction of the IIA or IIB theories.  In this scheme, the vertical
dimensional reduction of the brane-tension relation between the
7-brane, the NS-NS string and R-R 5-brane would lead to the $D=9$
relation
%%%%%
\be T_\8 = \fft{1}{2\pi}\, T_\2^{\ns}\, T_\6^{\rr} \ .
\ee 
%%%%%
However, this could not actually arise within the framework of a 
standard Kaluza-Klein reduction, since there is no seven-brane in 
$D=9$ {\it massless} supergravity.  It is, however, nevertheless 
consistent to perform instead a generalised Scherk-Schwarz dimensional 
reduction, which gives rise to a massive supergravity in $D=9$ 
\cite{bergre}, within which the above brane-tension relation does 
hold.  Applying T-duality and oxidising back to $D=10$, one is led to 
expect that there should be a brane-tension relation
%%%%%
\be
T_\9 = \fft{1}{2\pi}\, T_\2\, T_\7\label{8brane}
\ee
%%%%%%%
in ten dimensions.  There is no eight-brane in massless type IIA
supergravity, but there is such a solution in massive type IIA
supergravity.  In the next sections, we shall show that the brane
tension relation (\ref{8brane}) does indeed hold within the framework
of the massive type II theory.

\section{Massive IIA supergravity}

\subsection{Doubled formalism for massive IIA supergravity}

   As originally formulated, the massive $N=2$ supergravity in ten
dimensions involved a fixed mass parameter $m$.  After a
transformation of variables, given in \cite{bergre}, its bosonic
sector can be described by the Lagrangian\footnote{Our notation and
conventions are different from those used in \cite{romans}; here we
use a convenient notation, using differential forms.  The Lagrangian
is written as a 10-form.  When there is no ambiguity, we often omit
the wedge-product symbol between differential forms in a product, for
example writing $A\wedge B$ as $A\, B$, and $A\wedge A$ as $(A)^2$,
{\it etc}. }
%%%%%
\bea
{\cal L} &=& R\, {*\oneone} -\ft12 {*d\phi}\wedge d\phi - \ft12
e^{\fft32\phi}\, {*F_\2}\wedge F_\2 - \ft12 e^{-\phi}\, 
{*F_\3}\wedge F_\3- \ft12 e^{\fft12\phi}\, {*F_\4}\wedge F_\4 \nn\\
&&\!\!
-\ft12 dA_\3\wedge dA_\3 \wedge A_\2 - \ft16 m\, dA_\3 \wedge (A_\2)^3 
 -\ft1{40} m^2\, (A_\2)^5 -\ft12 m^2\, e^{\fft52\phi}\, {*\oneone}\ ,
\label{romans1}
\eea
%%%%%
where the field strengths are given in terms of potentials by
%%%%%
\bea
F_\2 &=& dA_\1 + m\, A_\2\ ,\qquad F_\3 = dA_\2\ ,\nn\\
F_\4 &=& dA_\3 + A_\1\wedge dA_\2 + \ft12 m\, A_\2\wedge A_\2
\ .\label{romfields}
\eea
%%%%%

   The formulation where $m$ is a constant is an inappropriate one in
the context of string theory, where one wishes to describe sets of
D8-branes that can carry different values of the ``charge'' $m$.  One
can easily reformulate the Lagrangian (\ref{romans1}) so that $m$ is
treated as a spacetime-independent field, subject to the Bianchi
identity $dm=0$.  This Bianchi identity can be enforced by adding a
Lagrange multiplier term ${\cal L}_{LM}= m\, dA_\9$ to
(\ref{romans1}).

     In this section, we shall reformulate the massive IIA theory in a 
``doubled formalism,'' following the same ideas and procedures as 
those developed in \cite{cjlp2}, where they were applied to the usual 
massless theories of $D=11$ supergravity, type IIB supergravity, and 
their toroidal dimensional reductions.  The philosophy of the doubled 
formalism is essentially to recast the system of second-order 
differential equations of motion for the original potentials of the 
theory into a first-order form, by introducing a dual potential for 
every original one.  The ostensible doubling of the physical degrees 
of freedom that would result from this is removed by the imposition of 
algebraic constraints that equate the new ``doubled'' set of field 
strengths to the duals of the original field strengths.  In fact, 
these constraint equations actually encode the original system of 
field equations.

     The strategy used in \cite{cjlp2} for constructing the doubled
systems was first to obtain the system of field equations from the
original Lagrangian describing the theory, and then to show by a
systematic procedure that each equation could be reformulated in a
first-order form, by introducing an appropriate dual potential.  In
our present massive IIA example, we begin by considering the equation
of motion for the 3-form potential $A_\3$ that follows from
(\ref{romans1}), namely
%%%%%
\be
d(e^{\fft12\phi}\, {*F_\4}) + dA_\2\, dA_\3 + \ft12 m \, A_\2\, A_\2
\, dA_\2=0 \ .
\ee
%%%%%
We note that an overall exterior derivative can be extracted from this
equation, so that we may write it as $d[ e^{\fft12\phi}\, {*F_\4}
+ A_\2\, dA_\3 +\ft16 m\, (A_\2)^3]=0$.  This allows us to re-express the
equation of motion as
%%%%%
\be
e^{\fft12\phi}\, {*F_\4} \equiv F_\6 = dA_\5 -A_\2\, dA_\3 - \ft16 m\,
(A_\2)^3\ ,\label{f4doubled}
\ee
%%%%%
where the 5-form potential $A_\5$ dual to $A_\3$ has now been
introduced.  Next, we consider the equation of motion for $A_\1$,
which is $d(e^{\fft32\phi}\, {*F_\2}) + e^{\fft12\phi}\, {*F_\4}\,
dA_\2=0$.  Substituting the previously-derived result
(\ref{f4doubled}) into this, we can then remove the derivative from this
equation, introducing a new doubled potential $A_\7$.  Continuing this
process, we can rewrite the entire theory in a first-order form by
introducing an additional double potential for each of the original
fields (including the dilaton, but excluding the metric itself). 
Summarising the results, we obtain the following first-order system:
%%%%%
\bea
e^{\fft12\phi}\, {*F_\4} &\equiv& F_\6 = dA_\5 -A_\2\, dA_\3 -\ft16 m\,
(A_\2)^3\ ,\nn\\
e^{\fft32\phi}\, {*F_\2} &\equiv& F_\8= dA_\7 - A_\2\, dA_\5 + \ft12
A_\2\, A_\2 dA_\3 + \ft1{24} m\, (A_\2)^4\ ,\label{romansdoubled}\\ 
e^{-\phi}\, {*F_\3} &\equiv& F_\7=dA_\6 - (dA_\5 -A_\2\, dA_\3 -\ft16m\,
(A_\2)^3 \, )\, \cA_\1\nn\\
&& - \ft12 A_\3\, dA_\3 - m\, A_\7\ ,\nn\\
{*d\phi} &\equiv& F_\9^\phi=dA_\8^\phi 
- \ft54 m\, A_\9 +\ft12 m\, A_\2\, A_\7 
- \ft34 A_\1\, dA_\7 -\ft12 A_\2\, dA_\6\nn\\
&& - \ft14 A_\3\, dA_\5
 +\ft34 A_\1\, A_\2\, dA_\5 + \ft14 A_\2\, A_\3\, dA_\3\nn\\
&&- \ft38 A_\1\, A_\2\, A_\2\, dA_\3 - \ft1{32}m\, A_\1\, (A_\2)^4 
\ ,\nn\\
m\, e^{\fft52\phi}\, {*\oneone} &\equiv& F_\ten= dA_\9 -A_\2\, dA_\7 +
\ft12 A_\2\, A_\2\, dA_\5\nn\\
&& -\ft16 (A_\2)^3\, dA_\3 -\ft1{120} m\,
(A_\2)^5\ .\nn
\eea
%%%%%
In deriving the last equation, we have treated $m$ as a 
spacetime-dependent field, and derived its ``equation of motion'' by 
varying the Lagrangian with respect to $m$.

   For future reference, we note that among the gauge symmetries of the 
double theory is one with a 1-form gauge parameter $\lambda_\1$, 
under which the various potentials transform as follows:
%%%%%
\bea
&&\delta A_\1 = -m\, \l_\1\ ,\qquad \delta A_\2 = d\l_\1\ ,\qquad 
\delta A_\3 = -m\, \l_\1\, A_\2\ ,\nn\\
&& \delta A_\5 = \l_\1\, dA_\3 - \ft12 m\, A_\2\, A_\2\, \l_\1\ ,\qquad
\delta A_\6 = \ft12 m\, A_\2\, A_\3\, \l_\1\ ,\label{stuck}\\
&&\delta A_\7 = \l_\1\, dA_\5 -\ft16 m\, (A_\2)^3\, \l_\1\ ,\qquad
\delta A_\9 = \l_\1\, dA_\7 -\ft1{24} m\, (A_\2)^4\, \l_\1\ .\nn
\eea
%%%%%
If we had been treating $m$ as a constant parameter in the Lagrangian, 
this transformation would have had the interpretation of describing a 
St\"uckelberg symmetry, which would allow the field $A_\1$ to be set 
to zero, reflecting the fact that this field is eaten by $A_\2$ when 
it becomes massive.  However since we have changed to a viewpoint in 
which $m$ is a field in the theory, we can no longer interpret $\delta 
A_\1 = -m\, \l_\1$ as the inhomogeneous term in a St\"uckelberg 
symmetry; rather, it is just one of many terms in the generally 
non-linear full set of transformations (\ref{stuck}).

    In fact, it is worth remarking that the treatment of the field $m$
can be put on a more equal footing with the other fields if we adopt
the formal device of regarding the 0-form field strength $m$ as
arising from the exterior derivative of a $(-1)$-form:
%%%%%
\be
m = d A_\m1\ .
\ee
%%%%%
Having done this, a sequence of transformations under the other gauge 
symmetries of the theory allow us to move the exterior derivatives off 
the $A_\m1$ potentials in (\ref{stuck}), and instead onto the gauge 
parameters $\l_\1$.  Having done so, we can then replace the exact 
2-form $d\l_\1$ by the closed 2-form $\Lambda_\2$, putting the gauge 
transformation of the 2-form potential $A_\2$ on a par with the way we 
have described the gauge transformations for all the other potentials 
in the doubled formalism.  We find that the full set of gauge 
transformations then takes the form
%%%%%
\bea
&&\delta A\sst{(-1)} = \L\sst{(-1)}\ ,\qquad
\delta A_\1 = \L_\1-\L_\2\, A\sst{(-1)}\ ,\qquad
\delta A_\2 = \L_\2\ ,\nn\\
&&\delta A_\3 = \L_\3-\L_\2\, A\sst{(-1)}\, A_\2+\L_\1\, A_\2\ ,\nn\\
&&\delta A_\5 = \L_\5 + \ft12\L_\1\, (A_\2)^2+\L_\2\, A_\3-\ft12 \L_\2\,
A\sst{(-1)}\, (A_\2)^2\ ,\nn\\
&&\delta A_\6 = \L_\6-\L_\1\, A_\5+
\ft12\L_\1\, A_\2\, A_\3+\L_\2\, A\sst{(-1)}\, A_\5\nn\\
&&\qquad\qquad - \ft12\L_\2\, A\sst{(-1)}\,
A_\2\, A_\3-\ft12\L_\3\, A_\3-\L_\7\, A\sst{(-1)}\ ,\nn\\
&&\delta A_\7 = \L_\7+ \ft16\L_\1\, (A_\2)^3+\L_\2\, A_\5-
\ft16\L_\2\, A\sst{(-1)}\, (A_\2)^2\ ,\nn\\
&&\delta A_\8 = \L_\8+ \ft34\L_\1\, A_\7-\ft18\L_\1\, (A_\2)^2\, A_\3-
\ft12\L_\2\, A_\6-\ft34\L_\2\, A\sst{(-1)}\, A_\7\nn\\
&&\qquad\qquad +\ft18\L_\2\, A\sst{(-1)}\, A_\3\, (A_\2)^2+\ft14\L_\3\,
A_\5+ \ft54\L_\9\, A\sst{(-1)}\ , \nn\\
&&\delta A_\9 = \L_\9+ \ft1{24}\L_\1\, (A_\2)^4+\L_\2\, A_\7-
\ft1{24}\L_\2\, A\sst{(-1)}\, (A_\2)^4\ .\label{gaugetransall}
\eea

    From the definitions of the original field strengths in
(\ref{romfields}), and the doubled fields in (\ref{romansdoubled}), it
is easy to calculate the Bianchi identities for the full set of field
strengths.  We find
%%%%%
\bea
&&dF_\2 = m\, F_\3\ ,\qquad dF_\3 =0\ ,\qquad dF_\4 = F_\2 \wedge
F_\3\ ,\nn\\
&&dF_\6 = -F_\3 \wedge F_\4\ ,\qquad dF_\8 = -F_\3 \wedge F_\6\ ,
\nn\\
&& dF_\7 = -\ft12 F_\4\wedge F_\4 - m\, F_\8 - F_\2\wedge F_\6\ ,\nn\\
&&dF_\9^\phi = \ft54\, m\, F_\ten - \ft34 F_\2\wedge F_\8 - \ft12 
F_\3\wedge F_\7 - \ft14 F_\4 \wedge F_\6\ ,\nn\\
&&dF_\ten = 0\ , \qquad d\, m =0\ .\label{rombianchi}
\eea
%%%%%
Note that these are all bilinear relations ($m$ is viewed a 0-form
field strength here).  It is interesting to note that although $F_\ten$
is by definition a closed 10-form, since we are in ten dimensions, we
could nevertheless choose to consider the system of field strengths in 
(\ref{romfields}) and (\ref{romansdoubled}) as being defined in some
arbitrary dimension $D>10$.  In this case, we can simply calculate
$dF_\ten$ from the definition of $F_\ten$ in (\ref{romansdoubled}),
finding
%%%%%
\be
dF_\ten = -F_\3 \wedge F_\8\ .\label{bianchiextra}
\ee
%%%%%
Thus even though there was no {\it a priori} reason for it to do so,
the field $F_\ten$ satisfies a bilinear Bianchi identity in $D>10$.

   From the Bianchi identities, it is a simple matter to read off the
commutation relations for the generators associated with the various
fields.  To do this, we first, as in \cite{cjlp2}, define the
generalised field strength obtained by summing over products of all
field strengths multiplied by their associated generators:
%%%%%
\bea
{\cal G} &=& \ft12 d\phi\, H + m\, e^{\fft54\phi}\, Y 
+ e^{\fft34\phi}\, F_\2\, W_1 +
e^{-\fft12\phi}\, F_\3\, V^1 + e^{\fft14\phi}\, F_\4\, V \nn\\
&&+  e^{-\fft14\phi}\, F_\6\, \wtd V + e^{\fft12\phi}\, F_\7\, \wtd V_1 
+  e^{-\fft34\phi}\, F_\8\, \wtd W^1 + e^{-\fft54\phi}\, F_\ten\, \wtd Y + 
\ft12 F_\9^\phi\, \wtd H\ .\label{2amassiveg}
\eea
Here, in addition to the generators already introduced for the usual
type IIA theory in \cite{cjlp2}, we have the generators $Y$ and $\wtd
Y$ associated with the 0-form and 10-form fields $m$ and $F_\ten$
respectively.  Note that these are both fermionic in nature, since the
associated potentials are odd-degree forms.

     It was shown in \cite{cjlp2} that the equations of motion can be
derived by requiring that the generalised field strength
${\cal G}$ satisfy the Cartan-Maurer equation
%%%%%
\be
d{\cal G} = {\cal G} \wedge {\cal G}\ .
\ee
%%%%%
This requirement then gives a determination of the commutation
relations for the various generators.  Thus we find by comparing with
(\ref{rombianchi}) that the non-vanishing commutators are precisely
those found in \cite{cjlp2} for the usual massless type IIA theory,
and presented here in (\ref{2athcom}), together with some additional
ones resulting from the inclusion of the additional fields $m$ and
$F_\ten$.  We find that these extra commutators are
%%%%%
\bea 
{[} V^1, Y {]} = W_1\ ,&& \{ \wtd W^1, Y \} = -\wtd V_1\ ,\nn\\
{[} H, Y {]} = -\ft52\, Y\ ,&& \{ Y, \wtd Y \} = -\ft58\, \wtd H\ .
\label{d10newalg1} 
\eea
%%%%%
Note that if we consider the theory in a dimension $D>10$, so that
there is the additional non-trivial Bianchi identity for $dF_\ten$, in 
(\ref{bianchiextra}), we obtain one further non-vanishing commutator,
namely
%%%%%
\be
{[} V^1, \wtd W^1 {]} = - \wtd Y\ .\label{d10newalg2}
\ee
%%%%%

   Actually, there are also direct ways of deriving this commutator that do
not require the use of the ``dimensionally-extended'' Bianchi identity
(\ref{bianchiextra}).  For example, one can read it off from the
gauge transformations given in (\ref{gaugetransall}).
Alternatively, one can read it off from the fact that the
exponentiation $\v$ of the
superalgebra generators in this theory, analogous to (\ref{2bcoset}),
must give rise to the generalised field strength ${\cal G} = d\v\,
\v^{-1}$ given in (\ref{2amassiveg}).  In general, all three
procedures give identical conclusions about the commutation relations
in the superalgebra \cite{cjlp2}, but the method where one reads them off 
from the Bianchi identities degenerates in the case of a field strength
of degree $D$ in $D$ dimensions unless one ``dimensionally extends''
the spacetime to $D+1$ dimensions.

\subsection{Massive type IIA brane tensions}

      The additional non-trivial commutators in the massive type IIA
supergravity imply additional brane-tension relations, over and above
those arising in the massless IIA theory.  In particular, the
commutator (\ref{d10newalg2}) implies that
%%%%%
\be
T_\9 = \fft1{2\pi}\, T_\2\, T_\7\ .\label{8brane2}
\ee
%%%%%%
The existence of this relation is essential for type IIA/IIB
T-duality, as explained in section 4.2.  It should be emphasised again
that the above relation is derived within the framework of the massive
type IIA supergravity itself, without needing to invoke type IIA/IIB
T-duality.  The first line of (\ref{d10newalg1}) can be understood by
applying the jade rule to (\ref{d10newalg2}).  Acting with the jade
rule on the brane tension (\ref{8brane2}), we would obtain two more
relations that involve the brane tension of a (-2)-brane.  It is not
clear whether such BPS states exist.

\section*{Acknowledgments}

We are grateful to R. Dijkgraaf and G. Papadopoulos for useful discussions.  
H.L. and K.S.S. would like to thank Texas A\&M University for hospitality.

\section*{Appendices}
\addcontentsline{toc}{part}{Appendices}

\appendix

\section{Massive IIA/IIB T-duality}

\subsection{Reduction of massive IIA to $D=9$}

   In order to find the T-duality transformation between the type IIB
and the massive type IIA theories, it is necessary to reduce each of
them to $D=9$.  For the type IIB theory the reduction will be of of
the generalised Scherk-Schwarz type, whereas for the massive type IIA,
it will be a standard Kaluza-Klein reduction on a circle.  In this
appendix, we now perform the reduction of the massive IIA theory.  We
shall be interested in obtaining the full doubled system in $D=9$.
For the most part, this can be done by dimensionally reducing the
already-doubled system in $D=10$.  However, since we are unable to
double gravity itself, it follows that a reduction of the doubled
$D=10$ system will not of itself generate the doubled fields for the
Kaluza-Klein vector and the new Kaluza-Klein dilaton, which come from
the ten-dimensional metric upon dimensional reduction.  Thus for these
fields, it is necessary to perform a doubling after having reduced the
ten-dimensional theory to $D=9$.  For all other fields, however, one
can easily check that the nine-dimensional theory obtained by doubling
in $D=10$ and then reducing to $D=9$ is the same as the one obtained
by reducing the original theory from $D=10$ to $D=9$ and then doubling
in the lower dimension.  Here, since the algebra is somewhat lengthy,
we shall just present our results for the full set of nine-dimensional
fields.  Since the doubling of the Kaluza-Klein fields must be
performed in $D=9$, it is useful first to present the nine-dimensional
Lagrangian:
%%%%%
\bea
{\cal L}_9 &=& R\, {*\oneone} -\ft12 {*d\phi}\wedge d\phi - 
\ft12{*d\vp}\wedge d\vp -\ft12 e^{\fft32\phi +2\alpha\vp}\, 
{* F_{\2}}\wedge F_\2 \nn\\
&&- \ft12  e^{\fft32\phi -14\alpha\vp}\, 
{* F_{\1 1}}\wedge F_{\1 1} 
 - \ft12  e^{-\phi +4\alpha\vp}\,  {* F_{\3}}\wedge F_{\3}
 - \ft12  e^{-\phi -12\alpha\vp}\,  {* F_{\2 1}}\wedge F_{\2 1}\nn\\
&& - \ft12  e^{\fft12\phi +6\alpha\vp}\,  {* F_{\4}}\wedge F_{\4}
- \ft12  e^{\fft12 \phi -10\alpha\vp}\,  {* F_{\3 1}}\wedge F_{\3 1}
- \ft12  e^{16\alpha\vp}\,  {* \cF_{\2}}\wedge \cF_{\2} \nn\\
&& -\ft12 m^2\, e^{\fft52\phi - 2\alpha\vp}\, {*\oneone}
 -\ft12 A_{\1 1}\, dA_\3\, dA_\3 - A_\2\, dA_{\2 1}\, dA_\3\nn\\
&&-\ft16 m\, (A_\2)^3\, dA_{\2 1}
- \ft12 m\, A_{\1 1}\, (A_\2)^2 \, dA_\3 -
\ft18 m^2\, (A_\2)^4\, A_{\1 1}\ ,
\eea
%%%%%
where we have defined $\alpha=1/(4\sqrt7)$.

   Now, we present the results for the doubled system.  Firstly, we
list the ``original'' field strengths in $D=9$:
%%%%%
\bea
&&F_{\1 1} = dA_{\0 1} + m\, A_{\1 1}\ ,\qquad 
F_\2 = dA_\1 + m\, A_\2 + \cA_\1 \, F_{\1 1}\ ,\nn\\
&&F_{\2 1} = dA_{\1 1}\ ,\qquad 
F_\3 = dA_\2 - \cA_\1\, dA_{\1 1} \ ,\nn\\
&& F_{\3 1}= dA_{\2 1} - A_{\0 1}\, dA_\2 + A_\1\, dA_{\1 1} +
m\, A_{\1 1}\, A_\2\ ,\nn\\
&&F_\4 = dA_\3 + + A_\1\, dA_\2 +\ft12m\, (A_\2)^2 + 
     \cA_\1\, F_{\3 1}\ ,\nn\\
&& \cF_\2 = d\cA_\1\ ,\label{2ad9orig}
\eea
%%%%%
Next, we list the doubled field strengths:
%%%%%
\bea
F_{\5 1} &=& dA_{\4 1} - A_{\1 1}\, dA_\3 - A_\2\, dA_{\2 1} -\ft12
m\, (A_\2)^2\, A_{\1 1}\ ,\nn\\
F_\6 &=& dA_\5 - A_\2\, dA_\3 -\ft16 m\, (A_\2)^3 + 
         \cA_\1\, F_{\5 1}\ ,\nn\\
F_{\6 1} &=& dA_{\5 1} - A_{\0 1}\, dA_\5 - A_\1\, dA_{\4 1} -
A_{\2 1}\, dA_\3 + A_{\0 1}\, A_\2\, dA_\3 
+ A_\1\, A_{\1 1}\, dA_\3 \nn\\
&&
+ A_\1\, A_\2\, dA_{\2 1} -m\, A_{\6 1} + \ft12m\, A_\1\, A_{\1
1}\,(A_\2)^2 
+\ft16 m\, A_{\0 1}\, (A_\2)^3\ ,\nn\\
F_\7 &=& dA_\6 -\cA_\1\, F_{\6 1} 
- A_\1\, dA_\5 -\ft12 A_\3\, dA_\3 + A_\1\, A_\2\,
dA_\3 -m\, A_\7\nn\\
&& + \ft16 m\, A_\1\, (A_\2)^3\ ,\nn\\
F_{\7 1} &=& dA_{\6 1} - A_{\1 1}\, dA_\5 - A_\2\, dA_{\4 1} +
A_{\1 1}\, A_\2 \, dA_\3 + \ft12 (A_\2)^2\, dA_{\2 1}\nn\\
&& + \ft16 m\, A_{\1 1}\, (A_\2)^3\ ,\nn\\
F_\8 &=& dA_\7 +\cA_\1\, F_{\7 1}
- A_\2\, dA_\5 +\ft12 (A_\2)^2\, dA_\3 + \ft1{24} m\, 
(A_\2)^4\ ,\nn\\
F_{\9 1} &=& dA_\8 - A_{\1 1}\, dA_\7 - A_\2\, dA_{6 1} + 
A_{\1 1}\, A_\2\, dA_\5 + \ft12 (A_\2)^2\, dA_{\4 1} \nn\\
&&
- \ft12 A_{\1 1}\, (A_\2)^2\, dA_\3 -\ft16 (A_\2)^3\, dA_{\2 1} -
\ft1{24} m\, A_{\1 1}\, (A_\2)^4\ ,\nn\\
\cF_\7 &=& d\cA_\6 - A_{\0 1}\, dA_{\6 1} - A_{\1 1}\, dA_{\5 1} -
A_{\2 1}\, dA_{\4 1} + A_{\0 1}\, A_{\1 1}\, dA_\5 \nn\\
&& + A_{\0 1}\, A_\2\, dA_{\4 1} + A_{\1 1}\, A_{\2 1}\, 
dA_\3 - \ft12 A_{\2 1}\, A_\3\, dA_{\1 1} \nn\\
&&-A_{\0 1}\, A_{\1 1}\, A_\2\, dA_\3 - \ft12 A_{\0 1}\,
(A_\2)^2\, dA_{\2 1} + m\, A_{\1 1}\, A_{\6 1}\nn\\
&& - \ft16 m\, A_{\0 1} \, A_{\1 1}\, (A_\2)^3
\ .\label{2ad9doubled}
\eea
%%%%%
Note that $\cF_\7$ is the doubled field strength corresponding to the
dual of Kaluza-Klein field strength $\cF_\2$.  We are using a notation
where a subscript ``1'' that is not enclosed in parentheses indicates
the internal index associated with the $D=10$ to $D=9$ reduction step.
Finally, the doubled fields associated with the two dilatons $\phi$
and $\varphi$ are:
%%%%%
\bea
F_{\8 1}^\phi &=& dA_{\7 1}^\phi - \ft34 A_{\0 1}\, dA_\7 - \ft34
A_\1\, dA_{\6 1} + \ft12 A_{\1 1}\, dA_\6 - \ft12 A_\2\, dA_{\5 1}
\nn\\
&&- \ft14 A_{\2 1}\, dA_\5 - \ft14 A_\3\, dA_{\4 1}
+\ft34 A_{\0 1}\, A_\2\, dA_\5 + \ft34 A_\1\, A_{\1 1}\, dA_\5 \nn\\
&&+\ft34 A_\1\, A_\2\, dA_{\4 1} - \ft14 A_{\1 1}\, A_\3\, dA_\3 
+\ft14 A_\2\, A_{\2 1}\, dA_\3 \nn\\
&&+ \ft14 A_\2\, A_\3\, dA_{\2 1} -\ft38 A_{\0 1}\, (A_\2)^2\, dA_\3 - 
\ft34 A_\1\, A_{\1 1}\, A_\2\,  dA_\3\nn\\
&& - \ft38 A_\1\, (A_\2)^2\, dA_{\2 1} 
- \ft54 m\, A_\8 - \ft12 m\, A_{\1 1}\, A_\7 + \ft12 m A_\2\, A_{\6 1} 
\nn\\
&&- \ft1{32} m\, A_{\0 1}\, (A_\2)^4 - \ft18 m\, A_\1\, A_{\1 1}\,
(A_\2)^3\ ,\nn \\
&&\nn\\
F_\8^\vp &=& dA_\7^\vp - 
\alpha\, \Big\{ -8 \cA_\1\, d\cA_\6 + 7 A_{\0 1}\, dA_\7 - A_\1
\, dA_{\6 1} + 6 A_{\1 1}\, dA_\6\nn\\
&& + 2 A_\2\, dA_{\5 1} + 5 dA_{\2 1}\, dA_\5 - 3 dA_\3 
\, dA_{\4 1} - 7 A_{\0 1}\, A_\2\, dA_\5\nn\\
&& + 8 A_{\0 1}\, \cA_\1\, dA_{\6 1} 
+ A\1\, A_{\1 1}\, dA_\5 + A_\1\, A_\2\, dA_{\4 1} - 3 A_{\1 1}\,
A_\3\, dA_\3 \nn\\
&&- 8 A_{\1 1}\, \cA_\1\, dA_{\5 1} -2 A_\2\, A_{\2 1}\, dA_\3 + 
A_{\2 1}\, A_\3\, dA_\2 + 
8 A_{\2 1}\, \cA_\1\, dA_{\4 1}\nn\\
&& + 8 A_{\0 1}\, A_{\1 1}\, \cA_\1\, dA_\5+
\ft72 A_{\0 1}\, (A_\2)^2\, dA_\3 - 
8 A_{\0 1}\, A_\2\, \cA_\1\, dA_{\4 1}\nn\\
&& - A_\1\, A_{\1 1}\, A_\2\, dA_\3 -\ft12 A_\1\, (A_\2)^2\, dA_{\2 1}
+ 8 A_{\1 1}\, A_{\2 1}\, \cA_\1\, dA_\3\nn\\
&& - 4 A_{\2 1}\, A_\3\, \cA_\1\, dA_{\1 1} - 8 A_{\0 1}\, A_{\1 1}\,
A_\2\, \cA_\1\, dA_\3 +4 A_{\0 1}\, (A_\2)^2\, \cA_\1\, dA_{\2 1}
\nn\\
&&+ m \, A_\8 - 6m\, A_{\1 1}\, A_\7 - 2m A_\2\, A_{\6 1} +
8m\, A_{\1 1}\, \cA_\1\, A_{\6 1}\nn\\
&&
+\ft{7}{24} m\, A_{\0 1}\, (A_\2)^4 -\ft16 m\, A_\1\, A_{\1 1}\,
(A_\2)^3 + \ft43 m\, A_{\0 1}\, A_{\1 1}\, \cA_\1\, (A_\2)^3\Big\}\ , 
\eea
%%%%%
where again $\alpha=1/(4\sqrt7)$.

   Before presenting the full set of nine-dimensional field equations
in the doubled formalism, it is useful to present a general lemma for
the dimensional reduction of field strengths and their duals.  If we
start with a metric in $(D+1)$ dimensions, and perform a reduction on a
circle to $D$ dimensions, the metric ansatz will be
%%%%%
\be
d\hat s^2 = e^{-2\alpha\vp}\, ds^2 + e^{2(D-2)\alpha\vp}\, (dz+\cA_\1)^2\ ,
\ee
%%%%%
where $\alpha=[2(D-1)(D-2)]^{-1/2}$.  The dimensional reduction of an
$n$-form field strength $\hat F_\n$ will give
%%%%%
\be
\hat F_\n = F_\n + F_{\sst{(n-1)}}\wedge (dz+\cA_\1)\ .
\ee
%%%%%
Denoting the Hodge dual in $(D+1)$ dimensions by $\hat *$, and in $D$
dimensions by $*$, it is easy to show that the dimensional reduction
of the dual of $\hat F_\n$ is given by
%%%%%
\be
{\hat * \hat F_\n} = (-1)^n\, e^{2(n-1)\alpha\vp}\, {*F_\n}\wedge (dz+
\cA_1) + e^{-2(D-n)\alpha\vp}\, {*F_{\sst{(n-1)}}}\ .
\ee
%%%%%
Consequently, we find that the dimensional reduction of the single
$(D+1)$-dimensional equation $e^{a\phi}\, {\hat * \hat F_\n} = \hat
{\wtd F}_{\sst{(D+1-n)}}$ will in general give rise to the two
$D$-dimensional equations
%%%%%
\bea
e^{a\phi +2(n-1)\alpha\vp}\, {* F_\n} &=& (-1)^n\, \wtd F_{\sst{(D-n)}1}\
,\nn\\
e^{a\phi -2(D-n)\alpha\vp}\, {* F_{\sst{(n-1)}1}} &=&  \wtd
F_{\sst{(D+1-n)}}\ .\label{d9doubled}
\eea
%%%%%

    Note therefore that although we {\it defined} the signs of our
doubled field strengths in the massive ten-dimensional theory to be
such that $e^{a\phi}\, {\hat * \hat F_\n} = +\hat {\wtd
F}_{\sst{(10-n)}}$, we are taking the doubled fields in $D=9$ to be
precisely those obtained by dimensional reduction of the
ten-dimensional doubled fields.  Consequently, we will have certain
minus signs in the duality equations in the nine-dimensional theory,
whenever $n$ is odd, as indicated in the first line in
(\ref{d9doubled}).

    With these preliminaries, we now present the nine-dimensional
equations of motion for the reduced massive IIA theory:
%%%%%
\bea
&& e^{\fft12\phi +6\alpha\vp}\, {*F_\4} = F_{\5 1}\ ,\qquad
   e^{\fft12\phi -10\alpha\vp}\, {*F_{\3 1}} = F_{\6}\ ,\nn\\
&&e^{-\phi +4\alpha\vp}\, {*F_\3} = -F_{\6 1}\ ,\qquad
  e^{-\phi -12\alpha\vp}\, {*F_{\2 1}} = F_{\7}\ ,\nn\\
&&e^{\fft32\phi +2\alpha\vp}\, {*F_\2} = F_{\7 1}\ ,\qquad
  e^{\fft32\phi -14\alpha\vp}\, {*F_{\1 1}} = F_{\8}\ , \\
&& {*d\phi} = - F_\8^\phi\ ,\qquad
   m\, e^{\fft51\phi- 2\alpha\vp}\, {*\oneone} = F_{\9 1}\ ,\nn\\
&& e^{16\alpha\vp}\, {*\cF_\2} = \cF_\7\ ,\qquad
   {*d\vp} = F_\8^\vp\ .\nn
\eea
%%%%%
Note that the two equations in the final line are associated with the
doubling of the new fields $\cA_\1$ and $\vp$ that have emerged from
the metric under dimensional reduction.  Since these have not
descended from any doubled equations in the higher dimension we have
simply chosen our definitions of the associated doubled field
strengths $\cF_\7$ and $F_\8^\vp$ so that there are plus signs in
these equations of motion.

\subsection{Doubled formalism for type IIB supergravity}

   The doubled formalism for type IIB supergravity was worked out in
detail in \cite{cjlp2}.  Here, we shall just summarise the results.
We do, however, make one change to the formalism, in anticipation of
the fact that we shall subsequently be using it for describing
Scherk-Schwarz generalised reductions.  It is therefore convenient to
make appropriate field redefinitions prior to constructing the doubled
formalism, such that the axion $\chi$ is covered by a derivative
everywhere.  The Lagrangian describing the bosonic sector of type IIB
supergravity may thus be written as
%%%%%
\bea
{\cal L} &=& R\, {*\oneone} - \ft12 {*d\phi}\wedge d\phi - \ft12
e^{2\phi}\, {*d\chi}\wedge d\chi-\ft12 e^{-\phi}\, 
{*G_\3^\ns}\wedge G_\3^\ns -\ft12 
e^{\phi}\, {*G_\3^\rr}\wedge G_\3^\rr\nn\\
&&- \ft14 {*G_\5}\wedge G_\5   + \ft12 B_\4\, dB_\2^\ns \,
dB_\2^\rr + \ft12 B_\4\, dB_\2^\ns\, dB_\2^\ns\, d\chi\ ,\label{2blag}
\eea
%%%%%
where the various field strengths are defined by
%%%%%
\bea
G_\3^\ns &=& dB_\2^\ns\ ,\qquad G_\3^\rr = dB_\2^\rr +B_\2^\ns\, d\chi
\ ,\nn\\
G_\5 &=& dB_\4 + \ft12 B_\2^\ns\, dB_\2^\rr -\ft12 B_\2^\rr\, dB_\2^\ns +
\ft12 B_\2^\ns\, B_\2^\ns\, d\chi\ .\label{2bfields}
\eea
%%%%%
As described in \cite{bbo}, the self-duality of $G_\5$ is to be
imposed here after varying the Lagrangian (\ref{2blag}) to obtain the
equations of motion.  This can be done consistently, since the
equation of motion for $G_\5$ turns out to be $d{*G_\5}= dB_\2^\ns\,
dB_\2^\rr + dB_\2^\ns\, dB_\2^\ns\, d\chi$, and the right-hand side is
identical to the expression for the Bianchi identity for $G_\5$,
following from (\ref{2bfields}).

    Following steps analogous to those used for the massive IIA theory
in the previous section, and described in detail in \cite{cjlp2}, we
can now construct the doubled formalism for the type IIB theory,
effectively re-expressing the second-order equations of motion
following from (\ref{2blag}) in first-order form.  We find the
following:
%%%%%
\bea
{*G_\5} &\equiv& G_\5 = dB_\4 + \ft12 B_\2^\ns\, dB_\2^\rr -
\ft12 B_\2^\rr\, dB_\2^\ns + \ft12 B_\2^\ns\, B_\2^\ns\, d\chi\ ,\nn\\
e^{\phi}\, {*G_\3^\rr} &\equiv & G_\7^\rr = dB_\6^\rr - B_\2^\ns\, dB_\4
-\ft14 B_\2^\ns\, B_\2^\ns\, dB_\2^\rr - \ft16 B_\2^\ns\, B_\2^\ns\,
B_\2^\ns\, d\chi\ ,\nn\\
e^{-\phi}\, {*G_\3^\ns} &\equiv& G_\7^\ns = dB_\6^\ns -B_\6^\rr\, d\chi
+ B_\2^\rr\, dB_\4 -\ft14 B_\2^\rr\, B_\2^\rr\, dB_\2^\ns + \ft14
B_\2^\ns\, B_\2^\ns\, B_\2^\rr\, d\chi\ ,\nn\\
e^{2\phi}\, {*d\chi} &\equiv& G_\9 = dB_\8 - B_\2^\ns\, dB_\6^\rr +
\ft12 B_\2^\ns\, B_\2^\ns\, dB_\4 + \ft18 (B_\2^\ns)^3\, dB_\2^\rr\nn\\
&& \qquad\quad + \ft18 B_\2^\ns\, B_\2^\ns\, B_\2^\rr\, dB_\2^\ns +
\ft1{24} (B_\2^\ns)^4\, d\chi\ ,\nn\\
{*d\phi} &\equiv& G_\9^\phi = dB_\8^\phi - B_\8\, d\chi -\ft12
B_\2^\ns\, dB_\6^\ns  + \ft12 B_\2^\rr\, dB_\6^\rr \nn\\
&&\qquad\quad +\ft12 B_\2^\ns\, B_\6^\rr\, d\chi - \ft12 B_\2^\ns\,
B_\2^\rr\,dB_\4 - \ft1{12} (B_\2^\ns)^3\, B_\2^\rr\, d\chi\ .
\eea
%%%%%

\subsection{Scherk-Schwarz reduction of IIB to $D=9$}

    Just as for the massive IIA theory discussed in Appendix A.1, here
too we may perform a dimensional reduction to obtain the doubled
formalism for type IIB in $D=9$.  This time, in order to make contact
with the nine-dimensional massive IIA theory, we must make a
generalised Scherk-Schwarz type reduction, where the axion $\chi$ in
$D=10$ is reduced according to $\chi(x,z) =\chi(x) + m\, z$.  All
other fields will be reduced according to the usual $z$-independent
Kaluza-Klein scheme.  Again, for all except the new Kaluza-Klein
vector, which we denote by $\cB_\1$ here, and the new Kaluza-Klein
dilaton $\vp$, the doubled fields in $D=9$ can be obtained simply by
dimensionally reducing the doubled fields in $D=10$.  However, to
obtain the doubles of the two new Kaluza-Klein fields we need to
perform a doubling in $D=9$.  It is therefore useful to begin by
presenting the type IIB Lagrangian in $D=9$:
%%%%%
\bea
{\cal L}_9 &=& R\, {*\oneone} - \ft12 {*d\phi}\wedge d\phi - \ft12
{*d\vp \wedge d\vp} - \ft12 e^{2\phi}\, {*G_\1}\wedge G_\1 
-\ft14 e^{-8\alpha\vp}\, {*G_\4}\wedge G_\4 \nn\\
&&- \ft14 e^{8\alpha\vp}\, {*G_\5}
\wedge G_\5 - \ft12 e^{-\phi+4\alpha\vp}\, {*G_\3^\ns}\wedge G_\3^\ns
- \ft12 e^{-\phi-12\alpha\vp}\, {*G_\2^\ns}\wedge G_\2^\ns\nn\\
&& - \ft12 e^{\phi+4\alpha\vp}\, {*G_\3^\rr}\wedge G_\3^\rr
 - \ft12 e^{\phi-12\alpha\vp}\, {*G_\2^\rr}\wedge G_\2^\rr 
-\ft12 e^{16\alpha\vp}\, {*\cF_\2}\wedge \cF_\2 \label{2bd9lag}\\
&&- \ft12 m^2\, e^{2\phi -16\alpha\vp}\, {*\oneone}
+\ft12 B_\3\, dB_\2^\ns\, dB_\2^\rr - \ft12 B_\4\, dB_\1^\ns\,
dB_\2^\rr + \ft12 B_\4\, dB_\2^\ns\, dB_\1^\rr \nn\\
&&+ \ft12 (B_\4\, B_\1^\ns\, dB_\2^\ns - B_\4\, B_\2^\ns\, dB_\1^\ns+ 
B_\3\, B_\2^\ns\, dB_\2^\ns)\, d\chi + \ft12m\, B_\4\, B_\2^\ns\,
dB_\2^\ns\ .\nn
\eea
%%%%% 
This is obtained by performing the Scherk-Schwarz reduction on the
ten-dimensional Lagrangian (\ref{2blag}).  

     Our results for the doubled system of fields in the massive
nine-dimensional type IIB theory are as follows.  Firstly, the
``original'' fields in $D=9$ are:
%%%%%
\bea
&&G_\1 = d\chi - m \, \cB_\1\ , \qquad G_\2^\rr = dB_\1^\rr + m\,
B_\2^\ns - \cB_\1\, d\chi\ ,\nn\\
&&G_\2^\ns = dB_\1^\ns\ ,\qquad 
G_\3^\ns = dB_\2^\ns - \cB_\1\, dB_\1^\ns\ ,\nn\\
&&G_\3^\rr = dB_\2^\rr + B_\2^\ns\, d\chi - \cB_\1\, dB_\1^\rr -
B_\1^\ns\, \cB_\1\, d\chi - m\, B_\2^\ns\, \cB_\1\ ,\nn\\
&&G_\4 =dB_\3 - \ft12 B_\1^\ns\, dB_\2^\rr + \ft12 B_\2^\ns\,
dB_\1^\rr + \ft12 B_\1^\rr\, dB_\2^\ns - \ft12 B_\2^\rr\, dB_\1^\ns
\nn\\
&&\qquad\quad -
B_\1^\ns\, B_\2^\ns\, d\chi + \ft12 m\, (B_\2^\ns)^2\ ,\nn\\
&&G_\5 = dB_\4 + \ft12 B_\2^\ns\, dB_\2^\rr -\ft12 B_\2^\rr\,
dB_\2^\ns - \cB_\1\, dB_\3 -\ft12 B_\1^\ns\, \cB_\1\, dB_\2^\rr \nn\\
&&\qquad\quad 
+\ft12 (B_\2^\ns)^2\, d\chi + \ft12 B_\2^\ns\, \cB_\1\, dB_\1^\rr + 
\ft12 B_\1^\rr\, \cB_\1\, dB_\2^\ns +\ft12 B_\2^\rr\, \cB_\1\,
dB_\1^\ns \nn\\
&&\qquad\quad - B_\1^\ns\, B_\2^\ns\, \cB_\1\,
d\chi -\ft12 m\, (B_\2^\ns)^2\, \cB_\1\ ,\nn\\
&& \cF_\2 = d\cB_\1\ .
\eea
%%%%%
We have presented these fields in the same order as the corresponding
fields of the nine-dimensional massive IIA theory in (\ref{2ad9orig}).
Of course in this case we also have $G_\5$ classified as an
``original'' field, since we had already effectively doubled the
$G_\5$ field in $D=10$.  In the type IIA picture, the corresponding
field $F_{\5 1}$ appears among the list of doubled fields in
(\ref{2ad9doubled}).

   We find that the doubled fields in $D=9$ are as follows:
%%%%%
\bea
&&G_\6^\rr = dB_\5^\rr + B_\1^\ns\, dB_\4 - B_\2^\ns\, dB_\3 + \ft12
N_\1^\ns\, B_\2^\ns\, dB_\2^\rr - \ft14 (B_\2^\ns)^2\, dB_\1^\rr
\nn\\
&&\qquad\quad +\ft12 B_\1^\ns\, (B_\2^\ns)^2\, d\chi -\ft16 m\,
(B_\2^\ns)^3\ ,\nn\\
&&G_\6^\ns = dB_\5^\ns - B_\1^\rr\, dB_\4 + B_\2^\rr\, dB_\3 + 
B_\5^\rr\, d\chi + \ft12 B_\1^\rr\, B_\2^\rr \, dB_\2^\ns \nn\\
&&\qquad\quad -\ft14 (B_\2^\rr)^2\, dB_\1^\ns -
\ft12 B_\1^\ns\, B_\2^\ns\, B_\2^\rr\, d\chi - \ft14 (B_\2^\ns)^2\,
B_\1^\rr\, d\chi + \ft14 m\, (B_\2^\ns)^2\, B_\2^\rr\ ,\nn\\
&&G_\7^\ns = dB_\6^\ns + B_\2^\rr\, dB_\4 - B_\6^\rr\, d\chi -
\cB_\1\, dB_\5^\ns - B_\1^\rr\, \cB_\1\, dB_\4
- \ft14 (B_\2^\rr)^2\, dB_\2^\ns\nn\\
&&\qquad\quad - B_\2^\rr\, \cB_\1\, dB_\3 + B_\5^\rr\, \cB_\1\, d\chi
+ \ft14 (B_\2^\ns)^2\, B_\2^\rr\, d\chi + \ft12 B_\1^\rr\, B_\2^\rr\,
\cB_1\, dB_\2^\ns \nn\\
&&\qquad\quad +\ft14 (B_\2^\rr)^2\, \cB_\1\, dN_\1^\ns - \ft12 
B_\1^\ns\, B_\2^\ns\, B_\2^\rr\, \cB_\1\, d\chi\nn\\
&&\qquad\quad - \ft14 (B_\2^\ns)^2\, 
B_\1^\rr\, \cB_\1\, d\chi + m\, B_\6^\rr\, \cB_\1\ ,\nn\\
&&G_\7^\rr = dB_\6^\rr - B_\2^\ns\, dB_\4 - \cB_\1\, dB_\5^\rr +
B_\1^\ns\, \cB_\1\, dB_\4 - \ft14 (B_\2^\ns)^2\, dB_\2^\rr +
B_\2^\ns\, \cB_\1\, dB_\3 \nn\\
&&\qquad\quad +\ft12 B_\1^\ns\, B_\2^\ns\, \cB_\1\, dB_\2^\rr
-\ft16 (B_\2^\ns)^3\, d\chi + \ft14 (B_\2^\ns)^2\, \cB_\1\,
dB_\1^\rr\nn\\
&&\qquad\quad +\ft12 B_\1^\ns\, (B_\2^\ns)^2\, \cB_\1\, d\chi + \ft16 m\,
(B_\2^\ns)^3\, \cB_\1\ ,\nn\\
&&G_\8 = dB_\7 + B_\1\, dB_\6^\rr - B_\2^\ns\, dB_\5^\rr - B_\1^\ns\,
B_\2^\ns\, dB_\4 + \ft12 (B_\2^\ns)^2\, dB_\3\nn\\
&&\qquad\quad - \ft38 B_\1^\ns\,(B_\2^\ns)^2\, dB_\2^\rr
-\ft14 B_\1^\ns\, B_\2^\ns\, B_\2^\rr\, dB_\2^\ns 
- \ft18 (B_\2^\ns)^2\, B_\1^\rr\, dB_\2^\ns\nn\\
&&\qquad\quad + \ft18 (B_\2^\ns)^3\, dB_\1^\rr 
+ \ft18 (B_\2^\ns)^2\, B_\2^\rr\, dB_\1^\ns 
- \ft16 B_\1^\ns\, (B_\2^\ns)^3\, d\chi +\ft1{24} m\, 
(B_\2^\ns)^4\ ,\nn\\
&&G_\9 = dB_\8 - B_\2^\ns\, dB_\6^\rr + \ft12 (B_\2^\ns)^2\, dB_\4 +
\ft18 (B_\2^\ns)^3\, dB_\2^\rr + \ft18 (B_\2^\ns)^2\, B_\2^\rr \,
dB_\2^\ns\nn\\
&&\qquad\quad + \ft1{24} (B_\2^\ns)^4\, d\chi -\cB_\1\, G_\8\ ,\nn\\
&&\cF_\7 = d\cB_\6 - \ft12 B_\3\, dB_\3 - B_\1^\ns\, dB_\5^\ns -
B_\1^\rr\, dB_\5^\rr + B_\1^\ns\, B_\1^\rr\, dB_\4 -\ft12 B_\1^\ns\,
B_\2^\rr\, dB_\3 \nn\\
&&\qquad\quad - B_\1^\ns\, B_\5^\rr\, d\chi + \ft12 B_\2^\ns\,
B_\1^\rr\, dB_\3 + \ft14 B_\1^\ns\, B_\2^\ns\, B_\1^\rr\, dB_\2^\rr + 
\ft18 B_\1^\ns\, (B_\2^\rr)^2\, dB_\1^\ns \nn\\
&&\qquad\quad +\ft18 (B_\2^\ns)^2\, B_\1^\rr\, dB_\1^\rr + \ft14
B_\2^\ns\, B_\1^\rr\, B_\2^\rr\, dB_\1^\ns + \ft14 B_\1^\n2\,
(B_\2^\ns)^2\, B_\1^\rr\, d\chi \nn\\
&&\qquad\quad -m\, B_\7 + m\, B_\1^\ns\, B_\6^\rr -\ft18 m\,
B_\1^\ns\, (B_\2^\ns)^2\, B_\2^\rr + \ft1{24} m\, (B_\2^\ns)^3\,
B_\1^\rr \ .
\eea
%%%%%
  
     Finally, the doubled fields associated with the two dilatons
$\phi$ and $\vp$ turn out to be:
%%%%%
\bea
&&G_\8^\phi = dB_\7^\phi + B_\7\, d\chi + \ft12 B_\1^\ns\, dB_\6^\ns
-\ft12 B_\2^\ns\, dB_\5^\ns - \ft12 B_\1^\rr\, dB_\6^\rr + \ft12
B_\2^\rr\, dB_\5^\rr \nn\\
&&\qquad\quad +\ft12 B_\1^\ns\, B_\2^\rr\, dB_\4 -\ft12 B_\1^\ns\,
B_\6^\rr\, d\chi + \ft12 B_\2^\ns\, B_\1^\rr\, dB_\4 - \ft12
B_\2^\ns\, B_\2^\rr\, dB_\3\nn\\
&&\qquad\quad -\ft12 B_\2^\ns\, B_\5^\rr\, d\chi + \ft14 B_\1^\ns\,
(B_\2^\ns)^2\, B_\2^\rr\, d\chi + \ft1{12} (B_\2^\ns)^3\, B_\1^\rr\,
d\chi \nn\\
&&\qquad\quad -m\, B_\8 + \ft12 m\, B_\2^\ns\, B_\6^\rr -\ft1{12}
(B_\2^\ns)^3\, B_\2^\rr\ ,\nn\\ 
&&G_\8^\vp = dB_\7^\vp - 2 \alpha\, 
\Big\{2 B_\3\, dB_\4 + 3 B_\1^\ns\, dB_\6^\ns +
B_\2^\ns\, dB_\5^\ns + 3 B_\1^\rr\, dB_\6^\rr + B_\2^\rr\, dB_\5^\rr 
\nn\\
&&\qquad\quad -4 \cB_\1\, d\cB_\6
-2 B_\3\, \cB_\1\, dB_\3 + 2 B_\1^\ns\, B_\2^\rr\, dB_\4 
-3 B_\1^\ns\, B_\6^\rr\, d\chi - 4 B_\1^\ns\, \cB_\1\, dB_\5^\ns 
\nn\\
&&\qquad\quad -2 B_\2^\ns\, B_\1^\rr\, dB_\4
+ B_\2^\ns\, B_\5^\rr\, d\chi - 4 B_\1^\rr\, \cB_\1\,
dB_\5^\rr - 4 B_\1^\ns\, B_\1^\rr\, \cB_\1\, dB_\4 \nn\\
&&\qquad\quad- \ft14 B_\1^\ns\,(B_\2^\rr)^2\, dB_\2^\ns
-2 B_\1^\ns\, B_\2^\rr\, \cB_\1\, dB_\3
+ 4 B_\1^\ns\, B_\5^\rr\, \cB_\1\, d\chi\nn\\
&&\qquad\quad - \ft14
(B_\2^\ns)^2\, B_\1^\rr\, dB_\2^\rr - \ft14 (B_\2^\ns)^2\, B_2^\rr\,
dB_\1^\rr + 2 B_\2^\ns\, B_\1^\rr\, \cB_\1\, dB_\3 \nn\\
&&\qquad\quad -\ft14 (B_\2^\ns)^2\, B_\2^\rr\, dB_\1^\ns + \ft14
B_\1^\ns\, (B_\2^\ns)^2\, B_\2^\rr\, d\chi - B_\1^\ns\, B_\2^\ns\,
B_\1^\rr\, \cB_\1\, dB_\2^\rr\nn\\
&&\qquad\quad + \ft12 B_\1^\ns\, (B_\2^\rr)^2\, \cB_\1\, dB_\1^\ns 
 -\ft14 (B_\2^\ns)^3\, B_\1^\rr\, d\chi + \ft12 (B_\2^\ns)^2\,
B_\1^\rr\, \cB_\1\, dB_\1^\rr\nn\\
&&\qquad\quad + B_\2^\ns\, B_\1^\rr\, B_\2^\rr\,
\cB_\1\, dB_\1^\ns - B_\1^\ns\, (B_\2^\ns)^2\, B_\1^\rr\, \cB_\1\,
d\chi + 4m\, B_\8 \nn\\
&&\qquad\quad- 4m\, B_\7\, \cB_\1 -m\,  B_\2^\ns\, B_\6^\rr 
+ 4m\, B_\1^\ns\, B_\6^\rr\, \cB_\1 + \ft12 m\, (B_\2^\ns)^3\,
B_\2^\rr \nn\\
&&\qquad\quad -\ft12 m\, B_\1^\ns\, (B_\2^\ns)^2\, B_\2^\rr\, \cB_\1 
+ \ft16 m\, (B_\2^\ns)^3\, B_\1^\rr\, \cB_\1\Big\} \ .
\eea
%%%%%

    It follows that the nine-dimensional equations of motion may 
again be read off from the ten-dimensional ones, by using 
(\ref{d9doubled}).  Thus we find
%%%%%
\bea
&& e^{8\alpha\vp}\, {*G_\5} = -G_\4\ ,\qquad
   e^{-8\alpha\vp}\, {*G_\4} = G_\5\ ,\nn\\
&& e^{\phi 4\alpha\vp}\, {*G_\3^\rr} = - G_\6^\rr\ ,\qquad
   e^{\phi -12\alpha\vp}\, {*G_\2^\rr} =  G_\7^\rr\ ,\nn\\
&& e^{-\phi +4\alpha\vp}\, {*G_\3^\ns} = - G_\6^\ns\ ,\qquad
   e^{-\phi -12\alpha\vp}\, {*G_\2^\ns} = G_\7^\ns\ ,\nn\\ 
&& e^{2\phi}\, {*G_\1} = - G_\8\ ,\qquad
   m\, e^{2\phi - 16\alpha\vp}\, {*\oneone} = G_\9\ ,\\
&& {*d\phi} = - G_\8^\phi\ ,\nn\\
&& e^{16\alpha\vp}\, {*\cF_\2} = \cF_\7\ ,\qquad
 {*d\vp} = G_\8^\vp\ .\nn
\eea
%%%%%
Note that the two equations on the top line are actually equivalent.
As in the type IIA reduction, the equations in the final line
correspond to the new fields $\cB_\1$ and $\vp$ coming from the
dimensional reduction of the metric, and we have chosen the
conventions  for their doubled field strengths so that there are plus
signs in these two equations of motion.

\subsection{Massive IIA/IIB T-duality in $D=9$}

     Having obtained the doubled formalism for both the massive type
IIA and type IIB theories in $D=9$, it is straightforward, albeit
tedious, to verify that the two sets of equations of motion are the
same, after appropriate field redefinitions.  The T-duality between
massive type IIA and type IIB was proven in \cite{bergre}, making use
of the St\"uckelberg symmetry.  In this section, we shall present the
explicit T-duality transformation rules for the doubled formalism.  We
shall present these field transformation rules in two sets, namely the
R-R sector and NS-NS sector.  For the R-R sector, we find that the
expressions for the type IIB fields in terms of the type IIA fields
are:
%%%%%%%
\bea
\chi &=& -A_{\0 1}\ ,\qquad B_{\1}^\rr = A_\1\ ,\qquad
B_{\2}^\rr = -A_{\2 1} + A_{\0 1}\, A_\2 + A_\1\, A_{\1 1}\ ,\nn\\
%%%%%
B_\3 &=& A_\3 -\ft12 A_\1\, A\2 -\ft12 \cA_\1\, A_{\2 1} +
         \ft12 A_{\0 1}\, \cA_\1\, A_\2\ ,\nn\\
%%%%%%%
B_\4 &=& A_{\4 1} - \ft12 A_\2\, A_{\2 1} - \ft12 A_\1\, A_{\1 1}\,
        A_\2 \nn\\
&&+\ft12 A_{\1 1}\, \cA_\1\, A_{\2 1} -
  \ft12 A_\0\, A_{\1 1}\, \cA_\1\, A_\2\ ,\nn\\
%%%%%%%
B_\5^\rr &=& A_\5 - \ft14 A_\1\, (A_\2)^2 -\ft12 \cA_\1\, 
A_\2\, A_{\2 1} +\ft12 A_{\0 1}\, \cA_\1\, (A_\2)^2\ ,\nn\\
%%%%%%%
B_\6^\rr &=& A_{\6 1} - \ft14 (A_\2)^2\, A_{\2 1} +
\ft1{12} A_{\0 1}\, (A_\2)^3 - \ft14 A_\1\, A_{\1 1}\, (A_\2)^2
\nn\\
&& +\ft12 A_{\1 1}\, \cA_\1\, A_\2\, A_{\2 1} -
\ft12 A_\0\, A_{\1 1}\, \cA_\1\, (A_\2)^2\ ,\nn\\
%%%%%%%%
B_\7 &=& A_\7 -\ft18 A_\1\, (A_\2)^3 -\ft18 \cA_\1\, A_{\2 1}\, 
(A_\2)^2 + \ft18 A_{\0 1}\, \cA_\1\, (A_\2)^3\ ,\nn\\
%%%%%%%
B_\8 &=& A_\8 - \ft1{24} (A_\2)^3\, A_{\2 1} - 
\ft18 A_\1\, A_{\1 1} (A_\2)^3 +
\ft18 A_{\1 1}\, \cA_\1\, (A_\2)^2\, A_{\2 1}\nn\\
&& -\ft18 A_\0\, A_{\1 1}\, \cA_\1\, (A_\2)^3\ ,\label{rrtdual}
%%%%%%%%
\eea

        For the NS-NS sector, we find that the T-duality
transformations are given by
%%%%%%%
\bea
\cB_\1 &=& A_{\1 1}\ ,\qquad
B_\1^\ns = \cA_\1\ ,\qquad B_\2^\ns = A_\2 - A_{\1 1}\, \cA_\1
\ ,\nn\\ 
%%%%%%
B_\5^\ns &=& A_{\5 1} - A_{\0 1}\, A_\5 +\ft12 A_{\2 1}\, A_\3
-\ft12 A_\1\, A_\2\, A_{\2 1}\nn\\
&&  -\ft14 \cA_\1\, (A_{\2 1})^2 +\ft12 A_{\0 1}\, A_\1\, (A_\2)^2
+\ft12 A_{\0 1}\, \cA_\1\, A_\2\, A_{\2 1}\nn\\
&& -\ft14 (A_{\0 1})^2\, \cA_\1\, (A_\2)^2\ ,\nn\\
%%%%%%%
B_\6^\ns &=& \cA_\6 - A_{\0 1}\, A_{\6 1} -\ft12 A_{\1 1}\, A_{\2 1}\, 
A_\3 -\ft14 A_\2\, (A_{\2 1})^2\nn\\
&&-\ft12A_\1\, A_{\1 1}\, A_{\2}\, A_{\2 1} +
\ft14 A_{\1 1}\, \cA_\1\, (A_{\2 1})^2 +
\ft1{12} (A_{\0 1})^2\, (A_\2)^3\nn\\
&&+\ft12 A_{\0 1}\, A_\1\, A_{\1 1}\, (A_\2)^2 -
\ft12 A_{\0 1}\, A_{\1 1}\, \cA_\1\, A_\2\, A_{\2 1}\nn\\
&& + \ft14 (A_{\0 1})^2\, A_{\1 1}\, \cA_\1\, (A_\2)^2\ ,\nn\\
%%%%%%%%%%
\cB_\6 &=& A_\6 -\ft14 A_1\, A_\2\, A_\3 - \ft14 \cA_\1\, A_{\2 1}\,
A_\3 +\ft14 A_{\0 1}\, \cA_\1\, A_\2\, A_\3\nn\\
&&-\ft18 A_\1\, \cA_\1\, A_\2\, A_{\2 1} +
\ft18 A_{\0 1}\, A_\1\, \cA_\1\, (A_\2)^2\ ,\nn\\
%%%%%%%%
B_\7^\phi &=& \ft34 A_\7^\phi + \ft{\sqrt7}{4} A_\7^\varphi -
A_{\0 1}\, A_\7 +\ft1{16} A_\2\, A_{\2 1}\, A_\3\nn\\
&&-\ft18 A_\1\, (A_\2)^2\, A_{\2 1} - \ft18 A_\2\, (A_{\2 1})^2\, \cA_\1
+\ft18 A_{\0 1}\, A_\1\, (A_\2)^3\nn\\
&&+\ft14 A_{\0 1}\, \cA_\1\, (A_\2)^2\, A_{\2 1} -
\ft18 (A_{\0 1})^2\, \cA_\1\, (A_\2)^3\ ,\nn\\
%%%%%%%%
B_\7^\varphi &=& \ft{\sqrt7}4\, A_\7^\phi -\ft34 A_\7^\varphi
+\ft58 A_\2\, A_{\2 1}\, A_\3 + A_1\, A_{\1 1}\, A_\2\, A_\3\nn\\
&&-A_{\1 1}\, \cA_\1\, A_{\2 1}\, A_\3 +A_{\0 1}\, A_{\1 1}\, \cA_\1\,
A_\2\, A_\3\ ,\label{nsnstdual}
\eea 
%%%%%%
The relation between the dilatonic scalars in the two nine-dimensional
theories are given by
%%%%%%
\be
\pmatrix{\phi \cr \vp}_{\rm IIA} =
\pmatrix{\ft34 & -\ft{\sqrt7}{4} \cr
                                           -\ft{\sqrt7}{4} & -\ft34}
\pmatrix{\phi \cr \vp}_{\rm IIB} \equiv M\,
\pmatrix{\phi \cr \vp}_{\rm IIB}
\ .\label{dils}
\ee
%%%%%%
Note that we have $M^{-1} = M$.   The dimensional reduction of
the ten-dimensional string metric to $D=9$ is given by
%%%%%%%
\bea
ds_{\rm str}^2 &=& e^{\ft12\phi}\, ds_{10}^2 \nn\\
&=& e^{\ft12\phi}\, (e^{-\vp/(2\sqrt7)}\, ds_9^2 +
e^{\sqrt7\vp/2} \, (dz_2 + {\cal A})^2 ) \ ,
\eea
%%%%%%%%
where $ds_{10}^2$ and $ds_9^2$ are the Einstein-frame metrics in
$D=10$ and $D=9$.  The radius of the compactifying circle, measured
using the ten-dimensional string metric, is therefore given by
$R=e^{\ft14 \phi +\ft{\sqrt7}{4}\vp}$.  Note that the dilaton vector
$\{\ft14, \ft14 \sqrt 7\}$ defining the radius is an eigenvector of
$M$, with eigenvalue $-1$.  It follows that the radii $R_{\rm IIA}$
and $R_{\rm IIB}$ of the compactifying circles, measured using their
respective ten-dimensional string metrics, are related by $R_{\rm
IIA}=1/R_{\rm IIB}$.

      Note that all the T-duality transformations between the massive
type IIA and type IIB theories are independent of $m$.  In particular,
this means that the relations between the type IIA and type IIB fields
in nine-dimensions are the same whether one is looking at the massive theories 
or the massless ones.  Note also that the relations between the
original ``undoubled'' sets of fields do not involve any of the
extended ``doubled'' system, and so by restricting attention just to the
original undoubled fields in (\ref{rrtdual}) and (\ref{nsnstdual}),
one obtains the explicit field relations for the standard undoubled systems.

\end{document}